\documentclass[preprint,amsmath,amssymb,aps,]{revtex4-1}
\usepackage{comment}
\usepackage{graphicx}
\usepackage{dcolumn}
\usepackage{bm}
\usepackage{natbib}
\usepackage{dsfont}

\usepackage{color}

\begin{document}


\title{Bose-Einstein condensation and Berezinskii-Kosterlitz-Thouless transition in 2D Nonlinear
Schr\"{o}dinger model}



\author{Sergey Nazarenko}
\affiliation{Mathematics Institute, The University of Warwick, Coventry, CV4-7AL, United Kingdom}
\affiliation{Laboratoire SPHYNX, Service de Physique de l'\'Etat Condens\'e, DSM,
IRAMIS, CEA Saclay, CNRS URA 2464, 91191 Gif-sur-Yvette, France}

\author{Miguel Onorato}
\affiliation{Dipartimento di Fisica, Universit\`{a} di Torino, Via Pietro Giuria 1, 10125 Torino, Italy}
\affiliation{INFN, Sezione di Torino, Via Pietro Giuria 1, 10125 Torino, Italy}

\author{Davide Proment}
\email{davideproment[at]gmail.com}
\homepage{http://www.uea.ac.uk/~xne12yku/}
\affiliation{School of Mathematics, University of East Anglia, Norwich Research Park, Norwich, NR4 7TJ, United Kingdom}



\date{\today}

\begin{abstract}
We analyse the Bose-Einstein condensation process and the Berezinskii-Kosterlitz-Thouless phase transition within the Nonlinear Schr{\"o}dinger model and their interplay with wave turbulence theory.
By using numerical experiments we study how the condensate fraction and the first order correlation function behave with respect to the mass, the energy and the size of the system.
By relating the free-particle energy to the temperature, we are able to estimate the Berezinskii-Kosterlitz-Thouless transition temperature.
Below this transition we observe that for a fixed temperature the superfluid fraction appears to be size-independent, leading to a power-law dependence of the condensate fraction with respect to the system size.
\end{abstract}


\maketitle

\section{Introduction}

When the occupation of a single particle state - usually the lowest energy one - of a Bose system becomes macroscopically occupied,
a Bose-Einstein condensate (BEC)
is said to emerge \cite{pitaevskii2003bose}.
Usually this phenomenon manifests itself under a critical temperature  called Bose-Einstein condensation temperature.

In an infinite homogeneous  system the condensation depends strongly
on the dimensionality.
Indeed, in three dimensions the condensate is thermally stable while in less dimensions
thermal fluctuations are strong enough to upset the long-range order.
This is commonly known as Mermin-Wagner theorem \cite{MerminWagner}
and it is due to the large-scale Goldstone modes (phonons), which  have diverging infrared contributions to the particle density in one and two dimensions
\cite{Bogoliubov2009}.
However, the long-range order can be restored in a finite system as the thermal fluctuations
have an intrinsic cut-off at large scales \cite{pitaevskii2003bose}.
Moreover, in 2D systems of interacting bosons an additional fascinating phase transition takes place.  
It is known as Berezinskii-Kousterlits-Thouless (BKT) transition after the physicists who discovered it in the early 1970s, and it predicts that, below a certain critical temperature $T_{BKT}$, the field correlations become {``long''} - decaying as a power law as opposite to the exponential decay law at higher temperatures \cite{Berezinskii, KT, kosterlitz1974critical}.

In the last decade several experiments using rarefied quasi two-dimensional Bose gases have been performed to study such transitions.
Experimental setup examples used strongly harmonically trapped Bose gases \cite{PhysRevLett.95.190403, hadzibabic2006berezinskii, PhysRevLett.99.040402}, Josephson-coupled Bose-Einstein condensates \cite{PhysRevLett.99.030401}, and pancake-shaped gases \cite{PhysRevLett.102.170401} with eventual changes of the interaction strength \cite{hung2011observation} leading to also strong interaction regimes \cite{PhysRevLett.110.145302}.
Moreover, many theoretical and numerical works were published on this subject ranging from theory on interactions and condensation in two-dimensional gases \cite{PhysRevA.44.7439, PhysRevLett.84.2551}, to numerical simulations using Monte Carlo techniques \cite{Prokof2001, PhysRevA.76.013613, PhysRevLett.100.190402} and (semi-)classical field methods for trapped \cite{PhysRevLett.96.020404, PhysRevA.77.023618, PhysRevA.79.033626} and homogeneous \cite{PhysRevA.81.023623} systems.

There exist many links between processes in classical fluid turbulence and in wave turbulence to the  Bose-Einstein condensation
 phenomenon
\cite{nazarenko2011,DNPZ-bec, Zakharov2005203, Lvov2003333,Nazarenko20061, nazon2007, connaughton2005, during2009breakdown, Vladimirova, berloff2002ssn, Proment:2012rt, PhysRevE.83.066311, PhysRevA.86.013624, sun2012observation, Shukla}.
 It was understood that interacting Bose systems have features similar to fluids, vortices and waves, and that they can undergo a chaotic dynamics similar to turbulence. Then, condensation can be interpreted as an analogue of an inverse energy cascade in 2D fluid turbulence or  an   inverse  cascade of wave-action in wave turbulence where the dissipative mechanisms carrying the direct cascade may be thought as prototypes of the  evaporative cooling.

Wave properties were also exploited for qualitative derivations of the algebraic decay of correlations in 2D interacting boson systems (acoustic modes - Bogoliubov phonons) and in explaining the BKT transition (vortices). A recent review describing physical ideas and approaches in this area is given in \cite{Hadzibabic}. However, much of the physical picture remains elusive from the point of view of understanding fundamental wave and vortex dynamics responsible for these effects. 
A sophisticated renormalization group approach allows to formalise the statistical mechanics aspects of the problem and put them on a more solid theoretical footing \cite{RevModPhys.51.659, doi:10.1080/01418639708241138}, but it does not help clarifying the underlying wave and vortex  dynamics and turbulence.
Note that the dynamics of waves and vortices is interesting not only  in 2D systems but is also crucial in 3D where the interaction between vortices, sound and Kelvin waves has been studied to understand quantum turbulence, see for instance \cite{PhysRevB.72.172505}.

The physical picture which is usually painted when describing the BKT transition remains oversimplified and unrealistic. 
It is said that the BKT {transition occurs when} the state changes from a gas of free hydrodynamic vortices  at $T>T_{BKT}$ to a gas of tight vortex dipoles at  $T<T_{BKT}$. But to derive an exponential decay of the field correlations for $T>T_{BKT}$ one usually appeals to Rayleigh-Jeans distribution of a non-interacting Bose gas, which contradicts the picture of hydrodynamic vortices because the latter are strongly nonlinear structures. Similarly, to derive a power-law decay of the field correlations for $T<T_{BKT}$ one often  uses Rayleigh-Jeans distribution of Bogoliubov phonons abandoning any appeal to hydrodynamic vortex pairs. Adding to the confusion, the 
Bogoliubov phonons are considered to be disturbances of a hypothetical superfluid density rather than of a uniform condensate density (the former tends to a constant and the latter tends to zero in the infinite box limit). However, there is no clear definition of such a superfluid density and there is no physical picture why is it legitimate to consider acoustic waves (phonons) in a fluid with such a density. 

The present paper is an attempt by specialists in fluid and wave turbulence to understand physical processes accompanying the BKT transition and the condensation in finite 2D systems and, if possible, find analogies and interpretations of these processes within the conceptual framework of turbulence. 
We chose an approach of a direct numerical simulation (DNS) of a universal model which is a classical object for both turbulence (including wave turbulence, superfluid turbulence and optical turbulence \cite{DNPZ-bec, Vladimirova, sun2012observation}) and for the condensed matter theory of the BKT transition. Namely we will simulate the two-dimensional (2D) defocusing Nonlinear Schr\"{o}dinger (NLS) equation (\ref{eq:2dNLS}), also known as the Gross-Pitaevskii equation in the Bose-Einstein condensate community.

To keep our model simple, we will not use any trapping potential to mantain the system homogeneous and not introduce any stochastic forcing representing a thermal bath, allowing the particle interactions (even when they are weak) to do the job of driving the system to the equilibrium thermodynamic state. 
This aspect of our model makes it different from the previous works, eg.   \cite{PhysRevA.81.023623}.
Our system will be finite dimensional with a cut-off in the momentum space, and the finite values of the temperature and the chemical potential will be determined by the initial values of the energy and the number of particles in the system.
Similar 2D NLS system was recently computed in \cite{Shukla} where the main emphasis was on the weakly nonlinear regimes for which an interesting phenomenon of a self-induced cutoff was discovered. In the present paper, we will study in detail the strongly nonlinear regimes associated with the BKT transition and with onset of condensation.
Note that in the fluid dynamics community  numerical studies of Fourier  truncated 
fluid systems have also been conducted recently, eg. for the truncated Euler equation \cite{PhysRevLett.95.264502} 
and Burgers equation \cite{PhysRevE.84.016301}.

The paper is divided as follows. 
In Section \ref{sec:model} we present the NLS model explaining its physical examples and definitions.
Sections \ref{sec:wt-4w} and \ref{sec:wt-3w} describe how to apply the wave turbulence theory to the two weakly nonlinear regimes described by NLS: the one in   absence and the other in presence of a strong condensate fraction respectively.
Section \ref{sec:hydro} deals with the hydrodynamical formulation of the problem and Section \ref{sec:bkt} gives an introduction to BKT transition. 
Section \ref{sec:numerics} represents the main part of the work and contains all the numerical results.
Finally, in Section \ref{sec:conclusion} we discuss our results and draw the conclusions.

\section{The model and definitions \label{sec:model}}

Let us consider the non-dimensional 2D defocusing NLS equation 
\begin{equation}
\label{eq:2dNLS}
i \partial_t \psi(\mathbf{x}, t) +  \nabla^2 \psi(\mathbf{x}, t) -
 |\psi(\mathbf{x}, t)|^2 \psi(\mathbf{x}, t) = 0, 
\,\, \;\;\; \mathbf{x}\in \mathbb{R}^2,  \;
t \in \mathbb{R}, \;\; \mathbf{\psi}\in \mathbb{C}.
\end{equation}
Under certain assumptions, two physical systems can be described by this model: a (quasi) two-dimensional highly occupied Bose gas and light propagating in a self-defocusing photorefractive crystal.
In the following we present these two physical systems and introduce the model properties and definitions which will be useful in our work. 

\subsection{NLS modelling physical systems}

In the BEC community the three-dimensional mean-field model called Gross-Pitaevskii equation is widely used to describe the dynamics of a Bose-Einstein condensate in the limit of zero temperature \cite{pitaevskii2003bose}. 
This model, which is nothing but the 3D NLS equation, can also be used to describe a (quasi-)two-dimensional BEC in which a tight external potential is considered to confine the motion in two dimensions by freezing the third dimension.
Rigorously derived in the limit $ T \rightarrow 0 $,  the NLS equation turns out to be still valid at finite temperatures, provided that the expected occupation number of all quantum levels considered in the system is much greater than one \cite{PhysRevLett.87.160402}.
In this limit the quantum nature of the operators and the role of commutators become less important and a mean-field description is still possible.
Thus, under the important assumption that the occupation number is much greater than one and using relevant scalings (see for further details in Appendix \ref{app:NLS-BEC}) the NLS model (\ref{eq:2dNLS}) suitably describes a (quasi)-two-dimensional Bose gas.
It is important to stress that now the NLS equation mimics the dynamics of a gas that may or may not have a macroscopic fraction of the total particles being in the lowest energy mode (the condensate mode) and it exhibits a BKT transition when the energy density of the system, which is related to the temperature, is decreased.

The NLS model is also widely used in optics to describe propagation of light in a medium \cite{newell1990nonlinear, kivshar2003optical, boyd1992nonlinear}.
In this framework light propagates along a specified direction and disperses in its transverse plane, so that the NLS equation is inherently (spatially) two-dimensional and the propagation axis takes the role of the time axis.
Depending on the intensity of the electromagnetic field, various nonlinear terms should be included in the standard defocusing NLS model.
Wave turbulence and condensation in optical systems have widely been studied, see for instance \cite{Picozzi20125440} and references therein. 
Recently, light condensation have been attempted by Sun {\it et al.} \cite{sun2012observation} but some criticisms have been raised on the validity of their experimental procedure \cite{2013arXiv1307.5034S} and no other experiments have validate independently these results.

\subsection{NLS properties and definitions}

NLS model conserves the total number of particles
\begin{equation}
N=\int |\psi|^2 d\mathbf{x}
\end{equation}
and the total energy 
\begin{equation}
\label{eq:E2E4}
 E=E_2+E_4 , \;\;\; \hbox{where} \;\;\;  E_2= \int |\nabla\psi|^2 d\mathbf{x} \;\;\; \hbox{and} \;\;\; 
 E_4=\frac 1{2}\int |\psi|^4 d\mathbf{x} 
\end{equation}
 are the free-particle and the interaction energies, respectively.

Let the system be in a double-periodic square box with side $L$.
Define the Fourier transform,
\begin{equation}
\label{eq:fourier}
\hat \psi_{\bf k} = \frac 1{L^2} \int_{box} \psi ({\bf x}) \, 
e^{-i  {\bf k} \cdot {\bf x}} \, d{\bf x},
\end{equation}
so that
\begin{equation}
\label{eq:inv-fourier}
\psi ({\bf x}) = \sum_{\bf k} 
\hat \psi_{\bf k} \, 
e^{i  {\bf k} \cdot {\bf x}} ,
\end{equation}
where wave vectors ${\bf k}$ take values on a truncated 2D lattice,
$${\bf k} = \left(\frac{2\pi}{L} \, m_x , \frac{2\pi}{L} \, m_y\right), \;\;\; m_x, m_y =
-\frac{n}{2}+1, -\frac{n}{2}+2, -\frac{n}{2}+3, ..., \frac{n}{2}.$$
Note that truncation of the momentum (Fourier) space is essential for our considerations. Without such truncation we would face an ``ultraviolet catastrophe", i.e. the system would cool off to zero temperature no matter how much we would input energy in it initially \cite{0951-7715-5-3-005}. 
In real physical systems the role of such a momentum cut-off is played by the scale where the occupation number becomes of order one and the system is no longer described by the mean-field NLS model. 
Indeed, particles with higher momenta behave like a classical Boltzmann gas with a rapidly decaying Maxwellian distribution - hence an effective cut-off momentum.
In the following we will express this cut-off in the wave vector space as $ k_{\max} = \pi n/L $, being $ n $ the number of grid points in both the $ x $- and $ y $-directions of the lattice. 

The mean density is 
\begin{equation}
\label{eq:mean-rho}
\overline \rho  = \frac 1{L^2} \int_{box} \langle |\psi ({\bf x})|^2 \rangle\, 
 d{\bf x} = \sum_{\bf k} 
\langle |\hat \psi_{\bf k} |^2\rangle,
\end{equation}
Here, the angle bracket mean an ensemble average, whereas the overline means a double average - over the ensemble and over the box.
Of course, because of the conservation of $N$, the ensemble averaging 
in this formula is not necessary provided that each realisation 
has the same
number of particles (in this case $\overline \rho =N/L^2$). 
However, this averaging allows us to relate the density and the spectrum.
 
The spectrum is defined as follows:
\begin{equation}
\label{eq:spectrum}
n_{\bf k} = \frac {L^2}{(2 \pi)^2} 
\langle |\hat \psi_{\bf k} |^2\rangle \, .
\end{equation}
The first order correlation function is
\begin{equation}
\label{eq:g1}
g_1({\bf r}) = \langle \psi({\bf 0}) \psi^* ({\bf r}) \rangle =
\sum_{\bf k} 
\langle |\hat \psi_{\bf k} |^2\rangle e^{i  {\bf k} \cdot {\bf r}}
\quad \mbox{and} \quad \lim_{L
\to \infty} g_1({\bf r}) = \int n_{\bf k} e^{i  {\bf k} \cdot {\bf r}} \, d{\bf k}.
\end{equation}
Conversely, in the infinite box limit
\begin{equation}
\label{eq:nk-g1}
 n_{\bf k} =   \frac 1{(2 \pi)^2} \int  g_1({\bf r}) e^{-i  {\bf k} \cdot {\bf r}} \, d{\bf r}.
\end{equation}
Relations between the mean density, the first-order correlation and the spectrum are
given by
\begin{equation}
\label{eq:mean-rho-g1}
\overline \rho  = g_1({\bf 0})
\quad \mbox{and} \quad \lim_{L
\to \infty}  \overline \rho = \int n_{\bf k} \, d{\bf k}.
\end{equation}
The spectrum at the zero mode is related with
 the first-order correlation
function as follows,
\begin{equation}
\label{eq:n0-g1}
n_0 = n_{{\bf k} = {\bf 0}} = \frac 1{(2 \pi)^2} \int_{box} g_1({\bf r}) \, d{\bf r}.
\end{equation}

Given initial values of $ N $ and $ E $, the system relaxes to its equilibrium
state and may develop or not a macroscopic condensate fraction 
$$ C = \frac{n_0}{\sum_{\bf k}  n_{\bf k}} = \frac{ 4 \pi^2n_0}{L^2 g_1({\bf 0})} =   \rho_0 / \overline \rho, $$
where $ \rho_0  = \langle|\psi_0|^2\rangle = \langle|\hat \psi_{{\bf k}= {\bf 0}} |^2 \rangle $ is the density of the particles in the zero momentum
mode - the condensate density.
Let us represent $\psi$ as a sum of its box-averaged value and fluctuations,
$$
\psi = 
\psi_0 + \tilde \psi; \,\,\,\,\,\, 
\psi_0 = \frac 1{L^2} \int_{box} \psi(\mathbf{x}) \, d\mathbf{x} \, .
$$

Now consider the Penrose-Onsager definition of the condensate density
\begin{equation}
\label{eq:nc-g1}
\rho_c =  
{g_1({\bf \infty}) }.
\end{equation}
It is easy to see that 
\begin{equation}
\label{eq:nc-n0}
\rho_c =   \lim_{L\to \infty} \rho_0 
\end{equation}
i.e. $\rho_c$ is the condensate density in the infinite-box limit.

Finally, we define the healing length by using the mean density as
\begin{equation}
\xi = \frac{1}{\sqrt{\overline \rho}} \, .
\end{equation}
It is straightforward to observe that this is the characteristic length scale at which the linear and nonlinear 
terms of the NLS equation (\ref{eq:2dNLS}) become comparable.
As the mean density is obtained after averaging over the box, here we define $ \xi $ in 
a statistical sense.
Indeed, in the following we will analyse not only the more conventional cases where a 
localised perturbation is present on a uniform condensate background, but also conditions 
where perturbations are so many and so strong that the uniform condensate background can be neglected
\footnote{Note that this point is particularly important when comparing our defined
non-dimensional 2D healing length with the one usually defined in
the BEC community. 
First, one should recall that our definition is for a two-dimensional
NLS equation and so appropriate scalings should be used as shown in Appendix 
\ref{app:NLS-BEC}. 
Second, our definition depends on the total number of bosons in the system, not
only on the macroscopic number present in the condensate: we recover the usual definition 
only if localised perturbations are negligible with respect of the uniform condensate background.}.

\section{Wave turbulence in the absence of  a condensate \label{sec:wt-4w}}

For high temperatures, the Bose gas is weakly interacting.
In the other words, for the typical wavenumber
we have $k^2 \gg \overline \rho$
and, therefore the nonlinear term is small compared to the linear
term in the NLS equation (\ref{eq:2dNLS}).
Such a small nonlinearity condition, together with an infinite box
limit and the assumption of randomness, 
represent a standard setup of the wave turbulence (WT) approach, which allows
to derive a four-wave kinetic equation \cite{musher-nls} for the spectrum:
\begin{eqnarray}
\dot n_k &=& 4 \pi  \int
 n_{ { \bf k}_1} n_{ { \bf k}_2} n_{ { \bf k}_3} n_{ \bf k} \left[
\frac{1}{n_{ \bf k}} + \frac{1}{n_{ { \bf k}_3}} - \frac{1}{n_{ { \bf k}_1}} - \frac{1}{n_{ { \bf k}_2}} \right] \times
\nonumber \\
&&
\;\;\;\;\; \;\;\; \delta(\omega_{  k}+\omega_{ {  k}_3}-\omega_{ { k}_1}-\omega_{ { k}_2})
 \delta({ \bf k}+{ { \bf k}_3}-{ { \bf k}_1}-{ { \bf k}_2})
\, d { { \bf k}_1} d { { \bf k}_2}  d { { \bf k}_3}.
\label{ke-nls}
\end{eqnarray}
where
\begin{equation}
\label{eq:omega}
\omega_k = k^2
\end{equation}
is the dispersion relation for the wave frequency.

In this four-wave WT regime, the leading effect of the weak nonlinearity, appearing at the lower order with respect to
the energy transfers described by the kinetic equation (\ref{ke-nls}), is an upward nonlinear frequency shift \cite{zakharov41kst} of the
dispersion curve  (\ref{eq:omega}) by a $\bf k$ independent value
\begin{equation}
\label{eq:omegaNL}
\omega_{NL} = 2 \, \overline \rho.
\end{equation}
As we will see later, such a shift is easily detectable in numerical simulations, and its presence is a good indication
that the system is in the four-wave condensate-free WT regime. 
On the contrary, we shall see that in presence of 
a strong condensate the frequency shift value at small wavenumbers is twice smaller.

A lot of WT studies focus on power-law spectra of Kolmogorov-Zakharov type which
arise in forced and dissipated wave systems. However, there is no forcing or dissipation
in our finite-dimensional system, and a more relevant solution of the kinetic equation
 (\ref{ke-nls})
in the steady state is given by the Rayleigh-Jeans distribution 
\begin{equation}
\label{eq:rj}
n_{RJ}(\mathbf{k})=\frac{T}{4 \pi^2 ({k}^2+\mu)},
\end{equation}
where $ T $ and $ \mu $ are the temperature and the chemical potential of the system respectively.

The free-particle energy density results in
\begin{equation}
\epsilon_2(T, \mu) \equiv E_2/L^2=\int_{box} k^2 n_{RJ}(\mathbf{k})d\mathbf{k}=
 \frac {T k_{\max}^2 }{\pi^2}-\mu \overline \rho.
\label{eq:epsilon_Tmu}
\end{equation}   
Note that the interaction energy is small compared to the free-particle energy in this case,
so the free-particle energy is approximately equal to the total (conserved) energy.
For the mean particle density we can write,
\begin{equation}
\overline \rho(T, \mu)=\int_{box} n_{RJ}(\mathbf{k})d\mathbf{k}
\approx \int_{circle} n_{RJ}(\mathbf{k})d\mathbf{k} =\frac T{4 \pi}
 \ln \left(1+\frac{4 k_{\max}^2}{\pi \mu}\right).
\label{eq:rho-of-T-mu}
\end{equation}   
Here, to obtain the analytical expression we have replaced integration over the square box by 
integration over a circle of the same area, $ r \le 2 k_{\max} / \sqrt{\pi} $, which, although not absolutely precise, is quite accurate
as we will see later.

Taking the inverse Fourier transform of the Rayleigh-Jeans distribution 
(\ref{eq:rj}), we find asymptotically for large $r = |\mathbf{r}| $,
\begin{equation}
g_1(r) \approx \frac{\pi^{1/2} T}{ (2r)^{1/2} \mu^{1/4}} \, e^{-\mu^{1/2} r}, \;\;\;\; \hbox{for} \;\;\; 
r \gg \mu^{-1/2},
\label{eq:g1exp}
\end{equation}   
which is the well-known result about the exponential decay of correlations in an uncondensed 2D
weakly interacting (or non-interacting) Bose gas \cite{Popov, Hadzibabic}.

\section{Wave turbulence in presence of a strong condensate \label{sec:wt-3w}}

A different type of WT can be considered with a strong coherent condensate component
  $$\psi_0 =  \sqrt{\rho_0 } \; e^{-i \rho_0  t}, \;\; \rho_0 =\hbox{const}>0$$ (uniform in the physical space),
  and weak random disturbances
$\phi({\bf x},t)$ on the background of this condensate,
\begin{equation}
\psi({\bf x},t) = \psi_0 \, (1+ \phi({\bf x},t)), \;\;\;\; |\phi| \ll 1.
\label{psi_phi}
\end{equation}
To develop the weak WT closure in this case, one has to diagonalize
the linear part of the dynamical equation
 with respect to condensate perturbations $\phi$.
 Such a diagonalization procedure is called Bogoliubov transformation, which in the NLS case is
 as follows
\cite{DNPZ-bec,nazarenko2011,Zakharov2005203},
 \begin{equation}
a_{{\bf k}}  = \frac{\sqrt{ \rho_0}}{2}  \left[ \left({\omega_k^{1/2} \over k} - {ik \over \omega_k^{1/2}}
\right) \hat \phi_{{\bf k}} +
 \left({\omega_k^{1/2} \over k} + {ik \over \omega_k^{1/2}}
 \right) \hat \phi_{-{\bf k}}^* \right],
\label{bogolubov}
\end{equation}
or conversely
 \begin{equation}
\hat \phi_{{\bf k}}  = \frac 1 {2\sqrt{\rho_0}}  \left[ \left({i\omega_k^{1/2} \over k} + {k \over \omega_k^{1/2}}
\right)  a_{{\bf k}} -
 \left({i\omega_k^{1/2} \over k} - {k \over \omega_k^{1/2}}
 \right)   a_{-{\bf k}}^* \right],
\label{bogolubov1}
\end{equation}
where  $a_k$ are the new normal amplitudes, and
$\omega_k$ is the new frequency of the linear waves,
\begin{equation}
\omega_k = k \sqrt{k^2 + 2 \rho_0  },
\label{bog_disp}
\end{equation}
usually called the Bogoliubov dispersion relation \cite{Bogoliubov2009}.
Correspondingly, the  relevant spectrum now is
\begin{equation}
\label{eq:spectrum-b}
n_{\bf k} = \frac {L^2}{(2 \pi)^2} 
\langle |\hat a_{\bf k} |^2\rangle.
\end{equation}
For long waves,
$ k^2 \ll \rho_0 $, the dispersion relation corresponds to acoustic waves - phonons,
\begin{equation}
\omega_k = c_s \, k, \quad c_s = \sqrt{2 \rho_0 }.
\label{bog_disp-sound}
\end{equation}
 For short waves,
$ k^2 \gg \rho_0 $, the dispersion relation is the same as for free particles,
\begin{equation}
\omega_k = k^2 .
\label{bog_disp-sound}
\end{equation}

When the substitution (\ref{psi_phi}) is made into the NLS equation, it is straightforward to notice that, at the first order of the expansion,  the nonlinear term becomes quadratic with respect to condensate perturbations $ \phi $.
This means that the three-wave interactions of the perturbations are present since the form of their dispersion relation (\ref{bog_disp}) allows three-wave resonances, that is $ \mathbf{k}-\mathbf{k}_1-\mathbf{k}_2 = 0 $ and $ \omega-\omega_1-\omega_2 = 0 $ can be both satisfied for a set of $ (\mathbf{k}, \mathbf{k}_1, \mathbf{k}_2) $.
Thus,  WT on background
of a strong condensate is now described by a standard three-wave  kinetic equation,
\begin{equation}
\label{kineq-R}
	\dot n_k
=
	\int
 \hspace{-0.5mm}\left( \mathcal{R}_{{\bf k}_1 {\bf k}_2}^{{\bf k}}
 - \mathcal{R}_{{\bf k}_1 {\bf k}}^{{\bf k}_2} -
\mathcal{R}_{{\bf k} {\bf k}_2}^{{\bf k}_1}\right) d \textbf{k}_1 d \textbf{k}_2
\ ,
\end{equation}
where 
\begin{equation}
\label{eq:R}
	\mathcal{R}_{{\bf k}_1 {\bf k}_2}^{{\bf k}}
=
	 \left| V_{{\bf k}_1 {\bf k}_2}^{{\bf k}} \right|^2 
\delta(\omega_{  k}-\omega_{ { k}_1}-\omega_{ { k}_2})
 \delta({ \bf k}-{ { \bf k}_1}-{ { \bf k}_2})
\left(n_{{\bf k}_1} n_{{\bf k}_2}-n_{{\bf k}_2} n_{{\bf k}} -n_{{\bf k}} n_{{\bf k}_1} \right)
\ 
\end{equation}
with the following interaction coefficient \cite{DNPZ-bec,Zakharov2005203,nazarenko2011},
\begin{equation}
\label{V-on-cond}
V_{{\bf k}_1 {\bf k}_2}^{{\bf k}_3}
=\sqrt{\rho_0  \omega_{k_1} \omega_{k_2} \omega_{k_3}} \left[ \frac{6}{\sqrt{\alpha_{k_1} \alpha_{k_2} \alpha_{k_3}}}
+\frac 1{2} \left(\frac{{\bf k_1} \cdot {\bf k}_2}{k_1k_2 \alpha_{k_3}}
+\frac{{\bf k}_2 \cdot {\bf k}_3}{k_2k_3 \alpha_{k_1}}
+\frac{{\bf k_3} \cdot {\bf k}_1}{k_3 k_1 \alpha_{k_2}}
\right)\right],
\end{equation}
where $\alpha_k = 2 \rho_0  +k^2$.
Rayleigh-Jeans equilibrium distribution now becomes
 \begin{equation}
\label{eq:RJ-bog}
n_k = \frac T{4 \pi^2 \omega_k}.
\end{equation}

Note that even for short waves,
$ k^2 \gg \rho_0 $, the kinetic equation remains of the three-wave type
as long as the fluctuations remain weaker than the condensate,
$ |\phi| \ll 1.$ In this limit we get from 
    (\ref{bogolubov1})
 \begin{equation}
\hat \phi_{{\bf k}}  = \frac {(i + 1)} {2\sqrt{\rho_0}}  \left[   a_{{\bf k}} -
 i  a_{-{\bf k}}^* \right],
\label{bogolubov2}
\end{equation}
and assuming that waves with $\bf k$ and with $-{\bf k}$
have the same spectra but are independent of each other phases, 
 \begin{equation}
\label{eq:ak-frre-partic}
\langle |\hat a_{\bf k} |^2\rangle \approx \langle |\hat \psi_{\bf k} |^2\rangle.
\end{equation}
Thus, the Rayleigh-Jeans spectrum in this case is formally the same
ar the Rayleigh-Jeans spectrum for the four-wave (condensate free) system
given by formula (\ref{eq:rj}) with $\mu =0$.
This fact will be important for us to understand the relation between
the temperature $T$ and the free-particle energy density $\epsilon_2$.

For long waves,
$ k^2 \ll\rho_0 $,  assuming that waves with $\bf k$ and with $-{\bf k}$
have the same spectra but are independent of each other phases, we get from 
    (\ref{bogolubov1})
 \begin{equation}
\label{eq:ak-phonons}
\langle |\hat a_{\bf k} |^2\rangle \approx 
\frac{\sqrt{2} \, k}{\sqrt{\rho_0}}
\langle |\hat \psi_{\bf k} |^2\rangle.
\end{equation}
In this limit the  Rayleigh-Jeans distribution is
 \begin{equation}
\label{eq:ak-phonons1}
\langle |\hat \psi_{\bf k} |^2\rangle \propto
\frac {T}{k^2}.
\end{equation}
This expression implies that at any finite $T$ 
the integral of the number of particles is logarithmically divergent at
$k \to 0$ \cite{Bogoliubov2009}, which is a manifestation of absence of
condensation in 2D NLS in an infinite system at any finite $T$. 
It the other words, our assumption that the fluctuations are weak
compared to the condensate, $|\phi| = |\tilde \psi /\psi_0| \ll 1$, fails in the equilibrium state
in the infinite box limit, so that the Bogoliubov expansion around the condensate
used in this section is inapplicable.

\section{Hydrodynamic reformulation and power-law correlations
\label{sec:hydro}}

We saw in the previous section that the formal Bogoliubov expansion around a
uniform condensate wavefunction $\psi_0$ fails in the equilibrium state.
However, the way this expansion fails is quite revealing if we analyse the
condensate disturbance $\tilde \psi$ in the physical space.
For waves which are much longer than $1/\sqrt{\overline \rho}$
there is a
complete equivalence of the NLS equation with the compressible
hydrodynamics thanks to Madelung transformation
$$\psi = \sqrt{\rho} \, e^{i \, \theta}.$$
Interpreting $\rho$ and ${\bf u} =  2 \nabla \theta$ as a fluid density
and a fluid velocity respectively, we get
the following equations of conservation of the fluid mass and the momentum
respectively,
\begin{eqnarray}
\label{madel1}
\frac{\partial \rho}{\partial t} + \nabla \cdot (\rho {\bf u}) =0, \\
\label{madel2}
\frac{\partial \theta}{\partial t} +  (\nabla \theta )^2 + \rho  =0.
\end{eqnarray}
Let us suppose to define the density and phase fluctuations over the condensate as $ \tilde \rho = \rho - \rho_0 $ and  
$ \tilde \theta = \theta - \theta_0 = \theta +\rho_0 t$ respectively, so that the velocity fluctuation is $ \tilde \bf u = 2 \nabla \tilde \theta $.
For the weak acoustic waves around the uniform density ${\rho_0 }$, the density and the
velocity disturbances are related as
$$
\frac{|\tilde \rho|}{\rho_0 } = \frac{|\tilde {\bf u}|}{c_s} \equiv \frac{2 |\nabla \tilde {\theta}|}{\sqrt{2 \rho_0 }}
\sim \sqrt{2} \, \frac{k \, | \tilde {\theta}|}{\sqrt{ \rho_0 }}.
$$
Thus, for long waves, $ k^2 \ll\rho_0 $, the fluctuations of the phase are much greater
than the relative fluctuations of the density. 
Therefore, the violation of the condition $|\phi| = |\tilde \psi /\psi_0| \ll 1$ when $\tilde {\theta} \sim 1$ 
is caused by the fluctuations of the phase and not of the density.

On the other hand, a perturbation theory can be developed
directly from the hydrodynamic equations (\ref{madel1}) and (\ref{madel2}) around a uniform
state at rest $\rho = \overline \rho, \;\; {\bf u}=0$, in which case the
approximation of weak nonlinearity is valid when $| {\bf u}| \sim k |\tilde \theta| \ll c_s =  
\sqrt{2 \overline \rho}$, whereas changes in the phase itself can be of order one or even greater \cite{Popov}.
If the phase variations are strong, then even for nearly constant $\rho $
we have $\overline \rho \ne \rho_0$. This state can be called a {\em quasi-condensate} because topologically it is
equivalent to the true uniform condensate in a sense that there are no vortex-like phase defects in it.

Clearly, one can no longer use formulae (\ref{bogolubov}) and (\ref{bogolubov1}) in this case,
and the expression   (\ref{eq:ak-phonons1}) is no more valid. 
Instead, one can obtain the correct
relations from the energy written in terms of the hydrodynamic variables.
In the limit  $ k^2 \ll\overline \rho$ we have
\begin{equation}
\label{eq:hydro-H}
E = E_2+E_4  \approx  \int \rho (\nabla \theta )^2 \, d {\bf x} + \frac 1{2} \int \rho^2  \, d {\bf x}.
\end{equation}
Correspondingly, for the acoustic wave energy $ \tilde E = E-E_0 $, where $ E_0 $ is the energy of the 
quasi-condensate, we have in the leading order
\begin{equation}
\label{eq:hydro-H-lin}
\tilde E = \overline \rho \int  (\nabla \tilde \theta )^2 \, d {\bf x} + \frac 1{2} \int \tilde \rho^2  \, d {\bf x}.
\end{equation}
Here, the two terms have  equal means by the virial theorem, so 
$\tilde E = 2\, \overline \rho \int  \langle (\nabla \tilde \theta )^2\rangle \, d {\bf x}$ or in Fourier representation
$$\tilde E = 2\, \overline \rho \, L^2 \sum_{\bf k} k^2 \, \langle |\hat \theta_{\bf k} |^2\rangle. $$
Rayleigh-Jeans distribution is characterised by an energy equipartition in which each
vibrational (phonon) mode has energy $T$; this distribution becomes
\begin{equation}
\label{eq:rj-theta}
\langle |\hat \theta_{\bf k} |^2\rangle =
\frac {T}{2\, \overline \rho \, L^2\, k^2}.
\end{equation}
Note that this is the same expression as the one which could be obtained from (\ref{eq:RJ-bog}) 
in which   $a_{\bf k}$  is expressed using (\ref{bogolubov1}) and then  the 
result is expressed in terms of the phase using expansion
$$
\psi \approx \sqrt{\overline \rho} \, e^{i\, \theta} \approx \sqrt{\overline \rho} \, (1 +i\, \theta -  \theta^2 +...)
$$
However, now the result (\ref{eq:rj-theta}) goes beyond validity of such an expansion because
$\theta$ is no longer required to be small.

Let us now find the physical space correlation function $g_1({\bf r})$ for large
$r=|{\bf r}|$ taking into account
that the main contribution to it will come from long-wave phonons
with significant variations of $\theta$ but negligible variations of $\rho$.
\begin{equation}
\label{eq:g1-theta}
g_1({\bf r}) = \langle \psi({\bf 0}) \psi^* ({\bf r}) \rangle 
\approx
\overline \rho \, \langle e^{i (\theta({\bf 0}) - \theta({\bf r}))} \rangle =
\overline \rho \,e^{-\langle(\theta({\bf 0}) - \theta({\bf r}))^2\rangle/2},
\end{equation}
where we assumed $\theta$ to be a Gaussian random variable.
In Fourier space we have
$$
\langle(\theta({\bf 0}) - \theta({\bf r}))^2\rangle =
\sum _{\bf k} \left| 1-e^{i {\bf k} \cdot {\bf r}} \right|^2 \langle |\hat \theta_{\bf k} |^2\rangle
=\frac {2T}{\overline \rho L^2} \sum _{\bf k} \frac{[\sin ( {\bf k} \cdot {\bf r}/2)]^2}{k^2}.
$$
Replacing in the large box limit the sum by an integral evaluated over a circle with radius up to the
phonon range, $ \sim 1/a $, we have
$$
\langle(\theta({\bf 0}) - \theta({\bf r}))^2\rangle \approx
\frac {T}{ 2 \pi^2 \, \overline \rho} \int_0^{2\pi} \int _{0}^{1/a} \frac{[\sin ( {\bf k} \cdot {\bf r}/2)]^2}{k^2} \, d{\bf k}
\approx \frac {T}{ 2 \pi \, \overline \rho} \, \ln (r/a), \;\;\;\; \hbox{for}  \;\;\; r \gg a \sim \xi.
$$
Here $ a $ is the characteristic vortex core radius, which precisely indicates the crossover scale 
between the phonon range and the vortex range  disturbances - 
see eq. (\ref{eq:aVortexCore}) and its respective paragraph for details.
Substituting this expression into
equation (\ref{eq:g1-theta}), {we finally obtain}
\begin{equation}
\label{eq:g1-power}
g_1({ r}) = \overline \rho \, \left( \frac{a}{r} \right)^{1/(\lambda^2 \overline \rho)}, 
\;\;\;\; \hbox{for}  \;\;\; r \gg a \sim \xi.
\end{equation}
Here, we introduced the notation for the thermal de Broglie length,
$$
\lambda = \sqrt{\frac {4\pi}{T}}.
$$
The law of algebraic decay of the correlations (\ref{eq:g1-power}) is the central result
in the statistical theory of 2D Bose systems.

As we can see, it is sufficiently rigorous and fully consistent with WT: it is rigorously 
valid as long as the density perturbations (phonons) remain weaker than the 
condensate density, $\tilde \rho \ll \overline \rho$, which is true  for sufficiently
low temperature. Moreover, under the same conditions one can use the three-wave WT
kinetic equation (\ref{kineq-R}) to describe evolution out of equilibrium (with $\rho_0$
replaced by $\overline \rho$).

On the other hand, 
for higher temperatures the density fluctuations grow and eventually become of the same 
order as the mean density, $\tilde \rho \sim \overline \rho$.
Thus, even though the perturbation
expansion based on the hydrodynamic formulation goes further than the one
based on the original Bogoliubov assumptions, it still fails for  larger $T$.
The most intriguing development in the statistical theory of 2D Bose systems
is the claim that law (\ref{eq:g1-power})  is valid even in the infinite-box limit
when $\overline \rho$ is replaced by a (finite in the $L\to \infty$ limit) {\em
superfluid density} $\rho_s$ \cite{Hadzibabic}:
\begin{equation}
\label{eq:nc-g1}
 g_1(r)= \rho_s \left(\frac {a}{r} \right)^{\alpha}; \,\,\,\,\,
\alpha = \frac {1 }{\lambda^2 \rho_s }.
\end{equation}
The legitimacy of introducing such an effective density could be derived from the fact
that for large density fluctuations the kinetic energy, the first term in (\ref{eq:hydro-H}),
is not equal to its constant density representation, the first term in (\ref{eq:hydro-H-lin}):
rather it is greater than it via the Cauchy-Schwartz inequality. Such an imbalance could
be phenomenologically corrected by replacing the constant density 
$\overline \rho$ with a constant superfluid density $\rho_s < \overline \rho $:
\begin{equation}
\label{eq:hydro-H-rhos}
E_2 =   \int \rho (\nabla \theta )^2 \, d {\bf x} \approx  \rho_s \int  (\nabla \theta )^2 \, d {\bf x}
< \overline \rho \, \int  (\nabla \theta )^2 \, d {\bf x}.
\end{equation}
For our purposes, introduction of a model Hamiltonian as in eq. (\ref{eq:hydro-H-rhos}) can be viewed as a definition of the superfluid density $\rho_s$.

Let us use  law (\ref{eq:nc-g1}) to find a relation between $\rho_s$ and $ \rho_0$. Integrating expression (\ref{eq:nc-g1}), we have
\begin{equation}
\label{eq:rhos-g1}
\rho_0 =   
 \frac 1{L^2} \int_{box} g_1({\bf r}) \, d{\bf r}
=\frac {\rho_s}{L^2} \int_{box}
  \left(\frac {a}{r} \right)^\alpha
\, d{\bf r}
\approx
\frac {2 \pi  \rho_s }{L^2} \int_{circle}
\left(\frac {a}{r} \right)^\alpha
\, r \, d{ r}
=
\frac {2 \pi^{\alpha/2} \rho_s }{(2-\alpha)}
\left(\frac {a}{L} \right)^\alpha,
\end{equation}
where again we replaced integration over the square box by integration over the 
circle of the same area, $r \le L/\sqrt \pi$.
Suppose that in the limit $L\to \infty$ the value of $\rho_s$ approaches a finite  limit. Then, at fixed $E$ and $N$, the condensate density $ \rho_0$  tends to zero as $1/L^\alpha$. 
Note that eq. (\ref{eq:rhos-g1}) can be viewed as a special case of the Josephson's relation; a discussion may be found in \cite{PhysRevB.76.092502, Holzmann30012007}.

\section{BKT transition
\label{sec:bkt}}

The BKT transition results in a change from an exponential character of the correlation decay
given by expression (\ref{eq:g1exp}) to a power-law decay as in (\ref{eq:nc-g1})
when temperature drops below a critical value $T=T_{BKT} $.

First of all, without considering any physics one can claim that for the BKT transition
to happen the nonlinearity (interaction) must fail to be weak in the 2D NLS
system. Indeed, in weakly interacting
systems the Rayleigh-Jeans distribution 
(\ref{eq:rj}) is the only relevant equilibrium solution in the large box limit.
One can take the ratio of the interaction and the free-particle energy
densities, 
\begin{equation}
\sigma = \frac{\epsilon_4}{\epsilon_2}=\frac{\int \rho^2 \, d\mathbf{x}}{2 \int |\nabla \psi|^2 \, d\mathbf{x}} , 
\label{eq:sigma}
\end{equation}
as
a measure of the interaction strength. 
We will see that  $\sigma(\rho, \epsilon_2)$
is indeed size-independent and that the BKT transition
occurs almost {precisely} when $\sigma=0.5$, i.e. then the nonlinear contributions
are almost the same as the linear ones (recall that the contribution of
$\epsilon_4$ into the dynamical equation comes with weight 2 because
of presence of two $\psi^*$'s in $\epsilon_4$).

As we saw in the previous section, the power-law decay of correlations (\ref{eq:nc-g1})
at sufficiently low temperatures can be obtained systematically, although 
replacing $\overline \rho$ with $\rho_s$ when passing to the infinite-box limit
is the least rigorous part in this approach.
However, the most striking prediction of the BKT theory is the value of the exponent of 
the power law just below $T=T_{BKT} $:
\begin{equation}
\label{eq:transition-exp}
\alpha = \frac {1 }{\lambda^2 \rho_s }
= \frac {T_{BKT}}{4\pi \rho_s} =1/4.
\end{equation}
The standard physical argument leading to this prediction
considers hydrodynamic vortices of radius $\xi$ and single quantum of circulation $ \kappa=4\pi $ in our dimensionless units (see eg. \cite{Hadzibabic}). 
In a 2D  box of size $L$ the energy of a single quantum vortex is 
$ E_v \approx 2 \pi \rho_s \ln(L/a) $ and its associated entropy is $ S \approx 2 \ln(L/a) $, i.e. log
of the maximum number of vortices of core size $ a $ that can be packed into the box of size $L$.
Then the free energy of the system is
\begin{equation}
F=E_v-TS \approx \frac T{2}{(\rho_s\lambda^2-4)}\ln\left(\frac{L}{a}\right),
\end{equation}
showing a  change of sign at $ \rho_s\lambda^2 =4$, which corresponds to the
BKT transition temperature and appearance of the power law with exponent (\ref{eq:transition-exp}).
Below $T=T_{BKT} $  we have $ F>0 $ so that formation of a new vortex is energetically unfavourable.
It is usually said that tight vortex dipoles would be present in this case, with the number of
such dipoles decreasing when the system is cooled more.
Above $T=T_{BKT} $  we have $ F<0 $ so that formation of new vortices is energetically favourable
and we should expect proliferation of vortices which results in a state with a``gas" of ``free" (i.e. unbound)
vortices.
 
It is not hard to see that the standard physical argument reproduced above is oversimplified,
at least in the case of the 2D NLS system.
Firstly, NLS vortices can only be viewed as hydrodynamic if the distance between them
is much greater than their core which we shall see being order of the healing length $\xi$. This is not the case neither below
nor above $T_{BKT} $.
Moreover, the fact that at $T=T_{BKT} $ the linear and the nonlinear energies are equal means that at this temperature
the mean distance between the vortices is of order $\xi$, and that raising temperature further above $T_{BKT} $
make the system more and more weakly nonlinear. In this case, the vortices are very dissimilar to their hydrodynamic
counterparts - they are often called ghost vortices, because they do not have a dynamical significance being 
zeroes of a weakly nonlinear wave field whose spectrum is described by the four-wave kinetic equation 
 (\ref{ke-nls}). Later, we will see that also dipoles are not the most typical vortex structures below $T_{BKT} $
unless $T$ is significantly less than $T_{BKT} $. These vortices are also insignificant for the observed 
dynamical and statistical properties  below $T_{BKT} $: as we have seen in the previous section, the low-frequency
phonons are much more important. 

Keeping in mind these inconsistencies of the physical argument based on the hydrodynamical vortices
which applied to the 2D NLS system, it appears to be even more striking that, as we will see later, it
prediction for the critical exponent $\alpha =1/4$ appears to be accurate.
This prediction at the BKT transition point is however a standard and rigorous result in statistical mechanics obtained using for example the XY model, which belongs to the same class of universality of the 2D NLS \cite{kosterlitz1974critical}.

\section{Numerical results \label{sec:numerics}}

We have performed a series of direct numerical simulations (DNS)
of the 2D NLS equation in a doubly periodic square box
using a pseudo-spectral method. 
Our numerical code computes the integration in time using a split-step method, a standard technique for this model
\cite{Proment:2012rt}.
We have tried several
different values of the box size $L$.
In all our simulations, both wavevector components $k_x$ and $k_y$ have been truncated
at the maximum absolute value of
\begin{equation}
k_{\max} = \frac{n}{2}\frac{2\pi}{L}=\pi \, .
\end{equation}
This corresponds to an effective  physical grid with spacing 
\begin{equation}
\Delta x=\frac{2 \pi}{2 k_{\max} }=1,
\end{equation}
so that $n=L$.

Our goal was to study thermodynamic the equilibrium states, so there were no damping or forcing terms
in our DNS. 
The initial conditions were chosen so that in all simulations the {mean density} was kept the same, $ \overline \rho =1 $,
and the energy took a range of different initial values.
Initially we populated a fraction of high energy modes setting random phases and then we waited till a final equilibrium statistically steady state is reached driven only by the nonlinear interactions.

With this choice of the simulation parameters the healing length $\xi $, defined in a statistical sense, takes the value 
\begin{equation}
\xi = \frac{1}{\sqrt{\overline \rho}} = 1.
\end{equation}
Near its centre, the  vortex solution behaves as $\psi \approx 0.6(x+iy)$ using healing length units \cite{west}, so that the vortex core size $ a $ can be estimated as
\begin{equation}
a \approx \frac{\xi}{0.6} \approx 1.7 = 1.7 \, \Delta x.
\label{eq:aVortexCore}
\end{equation}
Thus, each vortex is resolved by approximately
$
\pi a^2 \approx 9
$ grid points.
For the corresponding wavenumber we have
\begin{equation}
\label{k-xi}
k_a = \frac{1}{a} \approx 0.6 \approx  k_{\max}/5.2,
\end{equation}
and in the low-temperature Bogoliubov regime we expect to see both the phonon range of scales, $k < k_a $, and  ``free particles" with
$k > k_a $. 
In fact, in this regime most of the particles will populate the condensate mode ($\mathbf{k}=0$) and the phonons, while most of the energy will be in the free particles (see further considerations and Fig. \ref{fig:BogoliubovSpectra}).

\subsection{The dispersion relation and the frequency shift}

\begin{figure}
\centering
\includegraphics[width=0.8\linewidth]{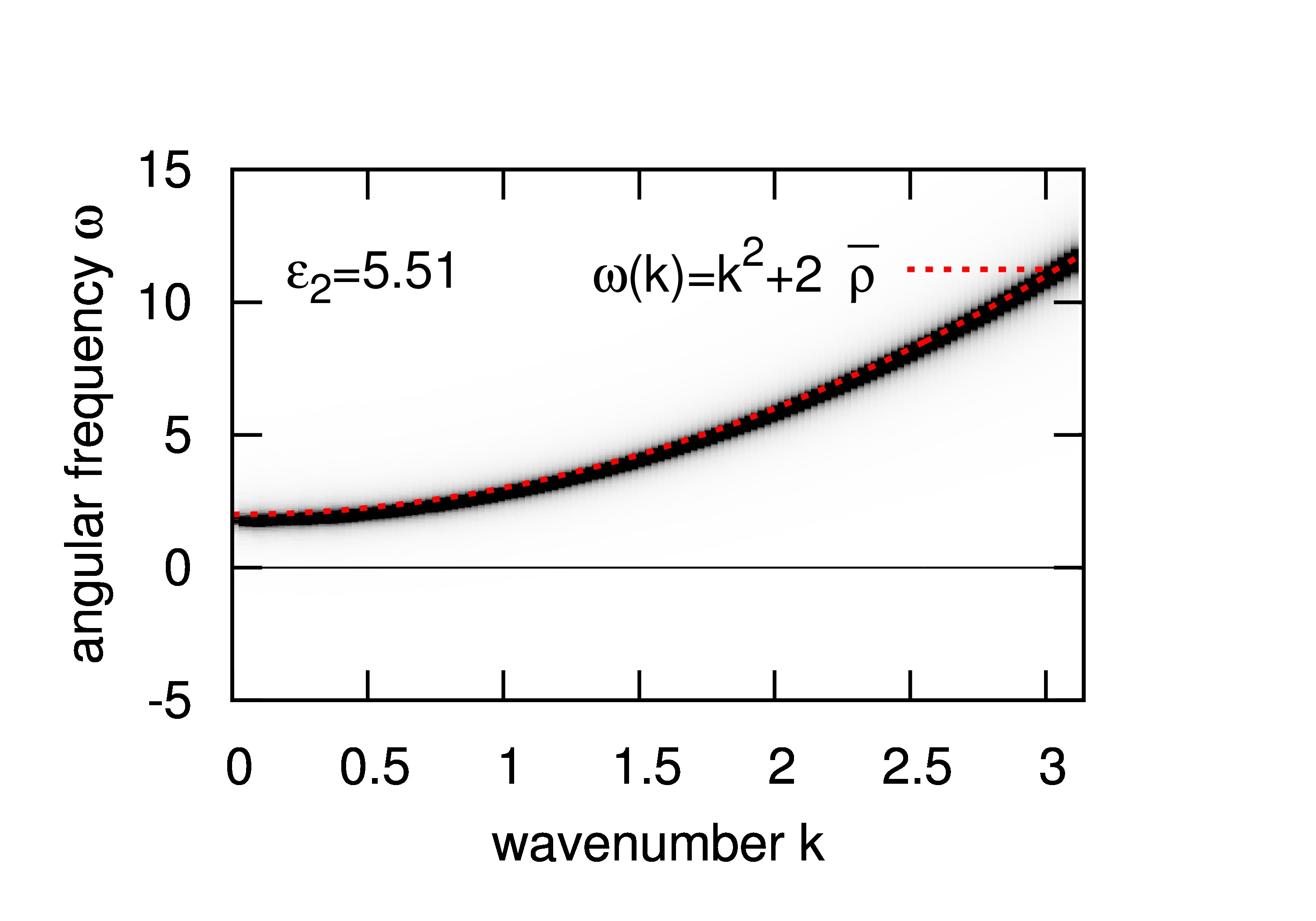}
\caption{
(Color online) Dark regions indicate the dispersion relation positive branch measured in a simulation with a weak (negligible) condensate. 
Results refer to the simulation having $ L=256 $ and free-particle energy density of $ \epsilon_2 =5.51 $.
The red dashed line shows the expected free-particle dispersion relation (\ref{eq:omega}) derived in the 4-wave interaction regime with the nonlinear frequency shift (\ref{eq:omegaNL}).
\label{fig:dr-1}}
\end{figure}

\begin{figure}
\centering
\includegraphics[width=0.8\linewidth]{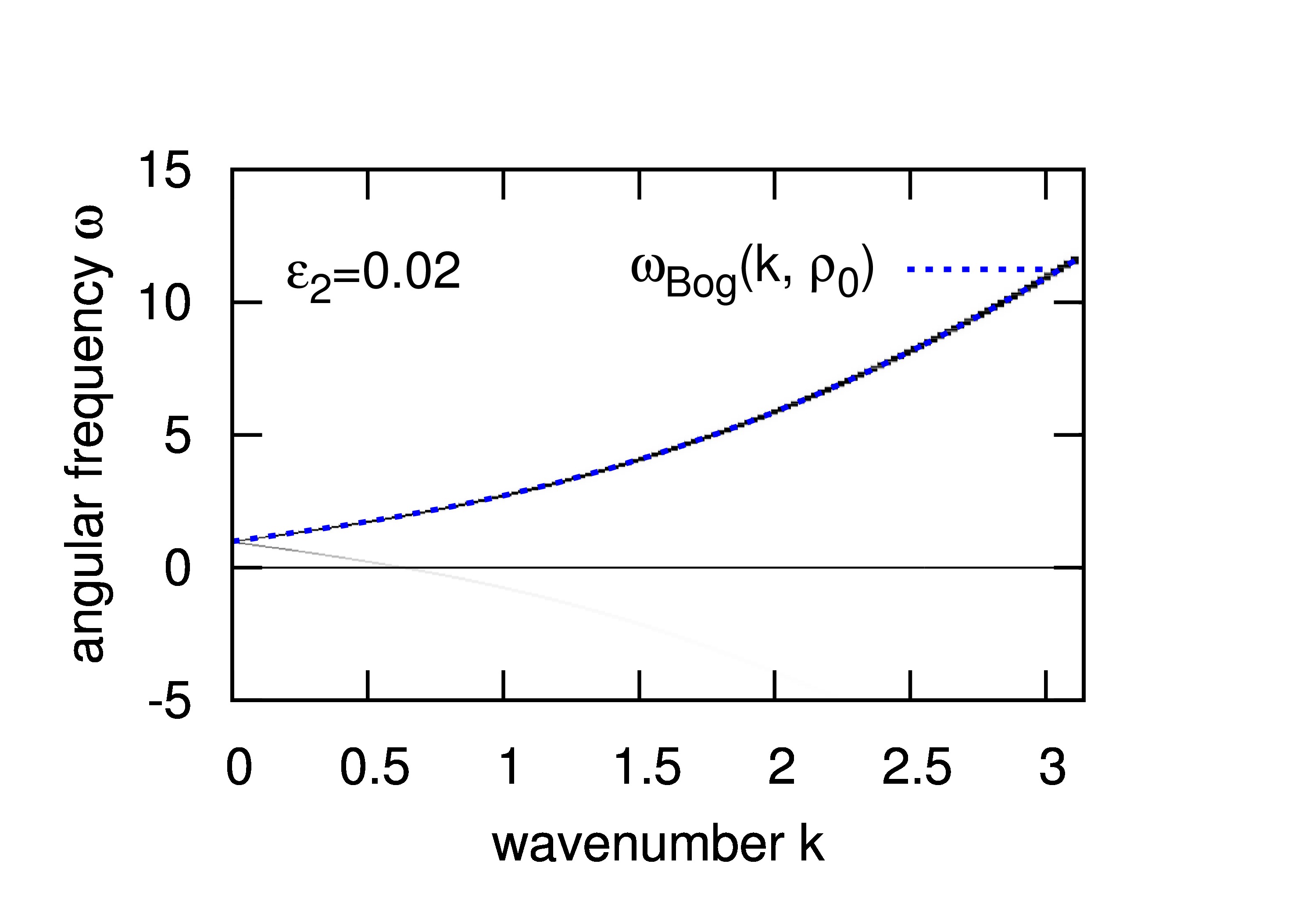}
\caption{
(Color online) Dark regions indicate the dispersion relation positive branch measured in a simulation with a strong condensate. 
Results refer to the simulation having $ L=256 $ and free-particle energy density of $ \epsilon_2 =0.02 $.
The blue dashed line shows the expected Bogoliubov dispersion relation (\ref{eq:Bogdisp}) derived in the 3-wave interaction regime.
\label{fig:dr-2}}
\end{figure}

First of all, let us examine the wave properties of the system and compare them with the theoretical
predictions for very high and very low temperatures.
For this, we perform a frequency Fourier transform of $\hat \psi_{\bf k}(t)$
 in the final steady state over a sufficiently long window 
of time and plot the resulting spectrum $\left|{\hat {{\hat {\psi}}}} ({{\bf k},\omega})\right|^2$ 
on a 2D plot $\omega$ vs $k = |{\bf k}|$.
Representative examples of such plots are shown in Fig.s \ref{fig:dr-1} and \ref{fig:dr-2}: a case of high $T$ on the former and a case of low $T$ on the latter, respectively.


 Our theoretical prediction for high temperatures is that the system is made of
 weakly interacting waves with the dispersion relation $\omega_k = k^2 + \omega_{NL}$
with the frequency up-shift $ \omega_{NL}$ as in (\ref{eq:omegaNL})
and the spectrum obeying the four-wave kinetic equation (\ref{ke-nls}).
On the $(k, \omega)$ plot such a weakly nonlinear system should manifest itself by a narrow
distribution around curve $\omega = \omega_k = k^2 + 2 \, \overline \rho$, as it  is
indeed seen in Fig. \ref{fig:dr-1}.
Note that the measured nonlinear shift $ \omega_{NL} $ is a little less than the theoretical $ 2 \, \overline \rho $, 
as this prediction is valid only in the limit $ E_2 \gg E_4 $.

For very low $T$, the theory predicts existence of two components - a strong
condensate rotating with frequency $\omega_0 = \rho_0 \approx \overline \rho$ and
weakly interacting phonons with Bogoliubov dispersion 
law (\ref{bog_disp}). 
Note that this dispersion is for the normal variable $a_{\bf k}$
whereas the Fourier transform of $\psi$ should be shifted by the condensate 
frequency $\omega_0 $ and should have two branches corresponding to $a_{\bf k}$
and $a_{\bf k}^*$ respectively,
\begin{equation} 
\omega_{Bog}(k)=\overline \rho \pm k \sqrt{k^2+2\, \overline \rho} \, .
\label{eq:Bogdisp}
\end{equation}
Once again, we can see a very good agreement with these predictions in
Fig. \ref{fig:dr-2}.
Indeed, we see both phonon branches as well as the shift due to the condensate rotating frequency.
Interestingly, we do not observe the horizontal line corresponding to a rotating  non-uniform condensate state as previously observed in forced and damped NLS simulations \cite{Nazarenko20061, nazon2007, Proment:2012rt}.
This is an indicator that  fully hydrodynamic vortices, making the condensate nonuniform and producing perturbations at $ k\neq0 $,  are  absent in the equilibrium state, whereas such vortices were present in previous simulations which focused on non-equilibrium states.

What happens at the intermediate temperatures?
This is the range where the classical wave turbulence based on weak nonlinearity
expansions fails and where
 strong turbulence takes place characterised by a complex interplay
of the vortex dynamics and strong acoustic pulsations. This is the range for which
power-law correlations and the  BKT
transition were predicted based on phenomenological physical arguments
presented in sections \ref{sec:hydro} and \ref{sec:bkt}.
There, the theory of Boboliubov type was extended beyond its formal limits
of applicability by introducing a hypothetical effective density - the superfluid 
density $\rho_s < \overline \rho$.  But we are now in a position to test
if replacing $\overline \rho$ by $\rho_s$ in the Bogoliubov theory, including 
 the dispersion relation, would be a good approximation to reality.
In particular we can examine the frequency shift $\omega_0 = \omega_{k=0}$
on the $(k, \omega)$ plots for the intermediate values of the temperature or, equivalently,
intermediate values of the energy.

\begin{figure}
\includegraphics[width=0.8\linewidth]{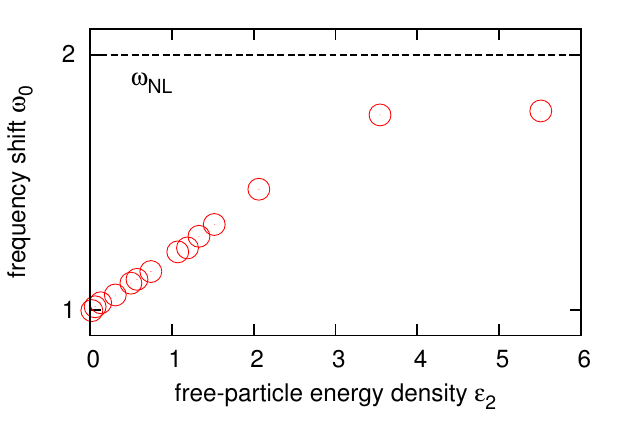}
\caption{
(Color online) Measured frequency shift as a function of the free-particle energy density $\epsilon_2$.
Results taken from simulations with $ L=256 $.
The frequency shift, estimated using the dispersion relation data, is expected to be closer to the pure condensate shift $ \rho_0 $ at low temperatures while to reach the limiting theoretical value $ \omega_{NL}=2\overline\rho $ (\ref{eq:omegaNL}) for high temperatures where condensate is negligible.
\label{omega-shift} }
\end{figure}

The result is shown in Fig. \ref{omega-shift} where we plot $\omega_0$ as a function of free-energy density $\epsilon_2$.
Here, we can see a monotonous increase of $\omega_0$ as $\epsilon_2$, and therefore $T$, increases
starting with the low-$T$ theoretical value of $ \overline \rho$ and aiming for another theoretical value
$2 \, \overline \rho$ on the high-$T$ side. This is in clear disagreement with a  naive replacement of 
$\overline \rho$ by $\rho_s$ in the Bogoliubov  
 the dispersion relation, because this would give us $\omega_0 =\rho_s$ which is a decreasing function of $T$.
Thus, strong turbulence in the intermediate range of turbulence surrounding the BKT transition 
cannot be fully understood as Bogoliubov phonons on the background of an effective superfluid density.

\subsection{Identification of $T$ with $\epsilon_2$}

First we will mention that for most experiments, except to the ones with the highest energy, we have assumed $T \approx \epsilon_2$; this will be justified later, see 
(\ref{TvsE}), but meanwhile, in the following subsections, this very useful fact will allow us to use $T$ instead of $\epsilon_2$.

\subsection{Strength of interactions}

\begin{figure}
\includegraphics[width=0.8\linewidth]{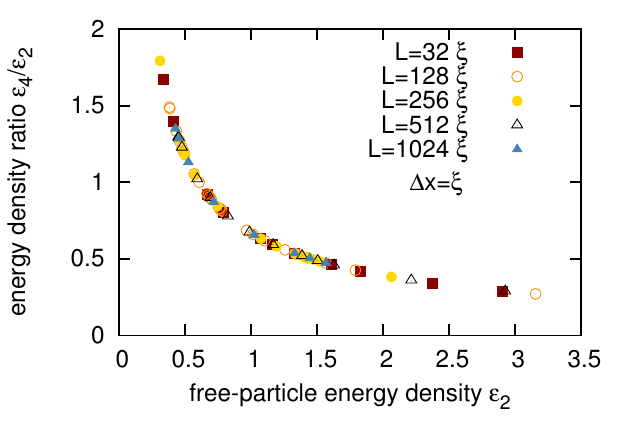}
\caption{
(Color online) Interaction energy density over the free-particle energy density $ \epsilon_{4}/\epsilon_2 $ as a function of the free-particle energy density $ \epsilon_{2} $ measured at final steady states in simulations having different system size.
Results show independence of the ratio with respect to the size of the system.
\label{fig:int-strength}}
\end{figure}

The interaction strength is a measure of the nonlinearity of the system. It can be quantified by the ratio of the interaction
energy to the energy of the free particles $ \sigma $ as defined in (\ref{eq:sigma}).
Figure \ref{fig:int-strength} shows the results for $\sigma$ as a function of  $ \epsilon_{2} $ in the steady state computed for 
$ \overline \rho = 1 $ and different values of $L$. We see that the data for the different values of $L$ collapse on the same
curve, which implies that for fixed $ \epsilon_{2} $ and $ \overline \rho = 1 $ the quantity $ \epsilon_{4} $ is $L$-independent.
This is an interesting and nontrivial result considering the fact that the many other quantities (eg. the condensate fraction, see
next section) are $L$-dependent.

\subsection{Condensate fraction}

\begin{figure}
\includegraphics[width=0.8\linewidth]{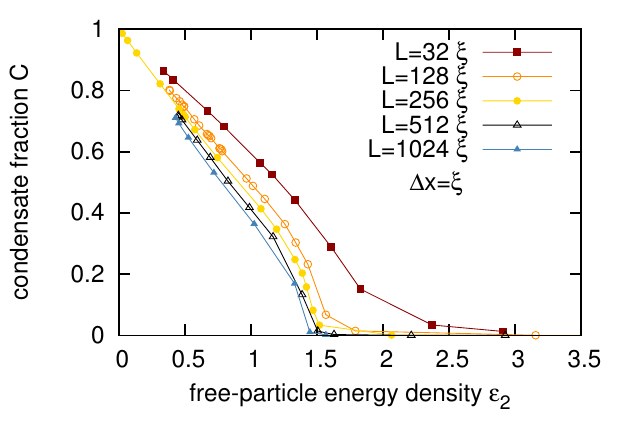}
\caption{
(Color online) Final steady state condensate fraction $ C=\rho_0/ \overline \rho $ as a function of the free-particle energy density $ \epsilon_2 $.
Different coloured symbols indicate different system sizes.
The condensate transition becomes shaper for larger sizes.
\label{fig:condensate-fraction}}
\end{figure}

\begin{figure}
\includegraphics[width=0.8\linewidth]{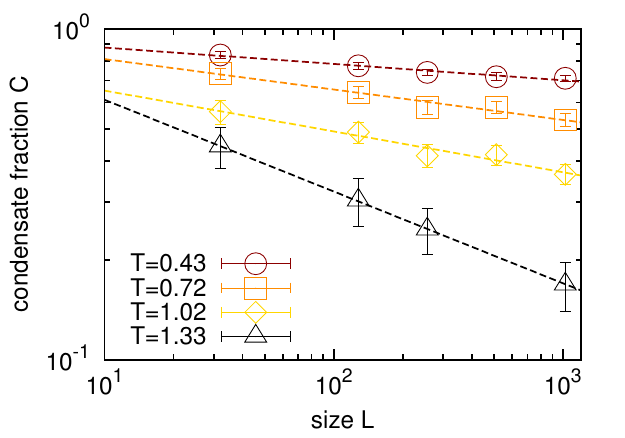}
\caption{
(Color online) Final steady state condensate fraction $ C=\rho_0/ \overline \rho $ as a function of the system size $ L $ for different temperatures in a log-log plot.
The straight dashed lines are the best fits obtained with the power law (\ref{eq:rhos-g1}).
Error bars indicate the time-averaged standard deviation of the condensate fraction fluctuation in time in the final steady states.
\label{fig:condensate-L}}
\end{figure}

Figure \ref{fig:condensate-fraction} shows the condensate fraction $C=\rho_0/\overline \rho $ in the final steady states as a function of free-energy density 
 $ \epsilon_{2} $ for systems having different sizes. 
One can see that $C$ depends on the box size $L$: as could be expected it is greater for smaller $L$.
Figure \ref{fig:condensate-L} shows the condensate fraction $C$ in the final steady states as a function of $L$ for different values of $ T \approx \epsilon_{2} $ below the condensation threshold.  
Note that $ \epsilon_2 $ cannot be set {\it a priori}, therefore we considered values within an error of 5\% in simulations having different L. 
We can see that $C$ is a decreasing  function of $L$ which approximately follows the theoretically predicted power law (\ref{eq:rhos-g1}).

Looking back at Fig. \ref{fig:condensate-fraction}, we can appreciate that for large $L$ the condensation threshold behaviour becomes more and more pronounced and a threshold value of
$ \epsilon_{2}$ (and therefore of $T$) tends to a limiting value $\epsilon_c \approx 1.4$ in the infinite-box limit. Of course, based on numerics alone, it is impossible
to establish if such a limiting value exists or the threshold tends to zero very slowly as a  function of $L$.
Argument in favour of the finite threshold limit $\epsilon_c $ was given before based on the view that for large $L$ the condensation is interaction-induced
and must occur when the  nonlinearity degree $\sigma$ reaches an order-one value.
As we see in Fig. \ref{fig:int-strength}, the value of  $\sigma$ is $\sim 0.5$ at $\epsilon_{2} = \epsilon_c \approx 1.4 $ 
which means that the linear and nonlinear terms in the NLS equation balance (recall again that the interaction energy contributes with factor 2 into the equation). 
In section \ref{sec:bkt} we presented a similar argument that the BKT transition should be expected when 
$\sigma$ reaches an order-one value. In fact, our numerics indicate that the condensation threshold temperature $T_c \approx \epsilon_c 
\approx 1.4$  is very close to the $T_{BKT}$
or may be even equal to it as we shall see in the following.

Figure \ref{fig:condensate-L} shows clearly that the condensate fraction $ C=\rho_0/ \overline \rho $ as a function of the box size $ L $ for different temperatures and we can see an excellent agreement with the predicted law (\ref{eq:rhos-g1}).
This result confirms the  claim that the value of the superfluid density $\rho_s$ is approximately $L$-independent. 

\subsection{Power-law correlations, superfluid density and BKT transition}

\begin{figure}
\includegraphics[width=0.8\linewidth]{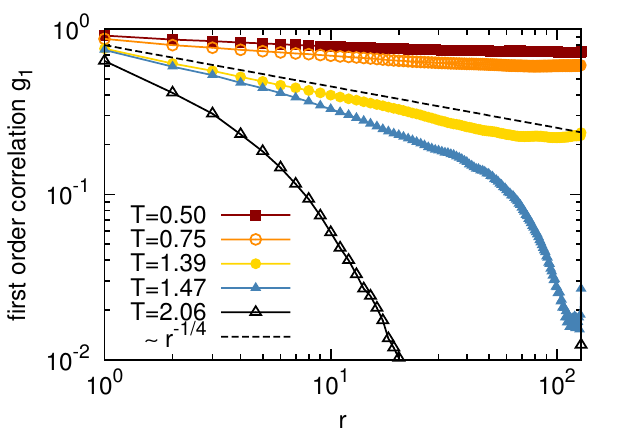}
\caption{ 
(Color online) Real part of the first-order correlation function of the field $ \psi $
for different temperatures as a function of an angle-averaged distance $ r $ in a log-log plot.
Results are taken from simulations having system size $ L=256 $.
The change from exponential to power-law behaviour characteristic to the BKT transition is clearly observed.
The power-law having slope $ \alpha=-1/4 $ at the transition temperature $ T_{BKT} $ is also shown.
\label{correlationLogLog}}
\end{figure}

At the BKT transition temperature $T_{BKT}$ 
the first-order correlation $ g_1(r)$ is predicted to change from
an exponential to a power-law decay, the former given in (\ref{eq:g1exp}) and the latter in (\ref{eq:nc-g1})
respectively.
In Fig. \ref{correlationLogLog}
we show a log-log plot of $ g_1(r)$ in simulations with $L=256$ 
and different values of the energy, i.e. corresponding to different temperatures.
We see that indeed a power-law behaviour appears 
for $T< T_{BKT} \approx \epsilon_2 = 
1.39$. Note that this value is very close to the condensation
threshold $ \epsilon_2 \approx 1.4$ for large $L$ shown in Fig. \ref{fig:condensate-fraction}.
As we mentioned above, the nonlinearity degree $\sigma$ has
value $\sim 0.5$ at this temperature, which corresponds to the balance of the
linear and nonlinear contributions in the NLS equation.
Ournumerical value for  $ T_{BKT} $  is different (although not by much) 
from the value $T_{BKT} = 4 \pi /\ln{760} \approx 1.89$
obtained in \cite{Prokof2001} by Monte Carlo simulations.
Probably this disagreement is due the different choice of the cut-off:
in our simulations the cut-off is kept fix while in \cite{Prokof2001} the cut-off
may vary as it is intrinsically set by the exponential tail of the Bose-Einstein distribution.
One can see in Fig. \ref{correlationLogLog} that the power-law exponent for temperatures
just below  $T_{BKT}$ agrees very well with the predicted value $\alpha =1/4$.
This is especially amazing considering non-rigorous character of the physical 
argument based on the picture of hydrodynamic vortices.
Indeed, as we mentioned in section \ref{sec:bkt} and as we will discuss in detail 
below in subsection \ref{subsec:vortices} the vortices in 2D NLS are never similar to their
hydrodynamic counterparts at thermodynamic equilibrium states of any temperature - most of the time
they lack nonlinearity to sustain themselves (ghost vortices) and they sporadically annihilate and get
created before they get a chance to move hydrodynamically.

We expect that at the BKT transition the relation 
$$
\lambda^2 \rho_s= \frac {4\pi \rho_s}{T_{BKT}} =4,
$$
holds. 
Using our numerical estimation $ T_{BKT} \approx 1.39 $ we obtain $ \rho_s \approx 0.44 $.
On the other hand we have from eq. (\ref{eq:rhos-g1})
\begin{equation}
\rho_s \approx
\frac {(2-\alpha)}{2 \pi^{\alpha/2} }
\left(\frac {L} {a}\right)^\alpha  \rho_0 \, .
\label{eq:estimrhos}
\end{equation}
We want to check the validity of such relation using for example the simulations with $ L =256 $.
Here at $ T=T_{BKT} \approx 1.39 $ we have measured numerically $ \rho_0 \approx 0.20 $. Substituting this value and $ a=1.7 $ into eq.  (\ref{eq:estimrhos}), we get
 $ \rho_s \approx 0.53 $ which  overestimates our previous result by about 20\%. 
Probably, the two main factors contributing to this deviation are the system periodicity and the square
box shape which are not taken into account during the calculation in (\ref{eq:rhos-g1}).



\begin{figure}
\includegraphics[width=0.8\linewidth]{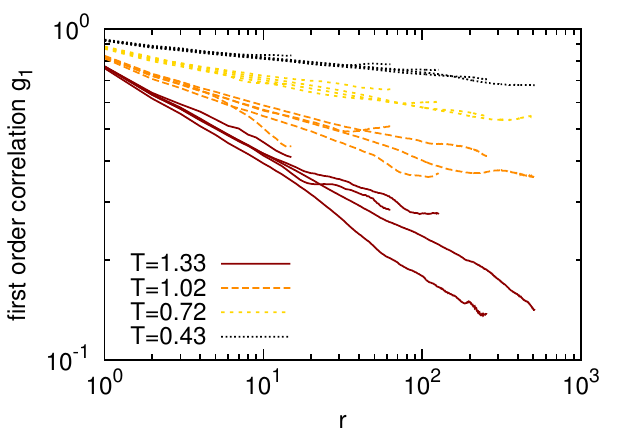}
\caption{
(Color online) Real part of the first-order correlation function of the field $ \psi $
for different temperatures below the estimated $ T_{BKT} $ as a function of an angle-averaged distance $ r $ in a log-log plot.
Same coloured lines refer to simulations having the same temperature but different system sizes: similar exponents in the power-laws are observed for the same temperature.
\label{correlationSize} }
\end{figure}
\begin{figure}
\includegraphics[width=0.8\linewidth]{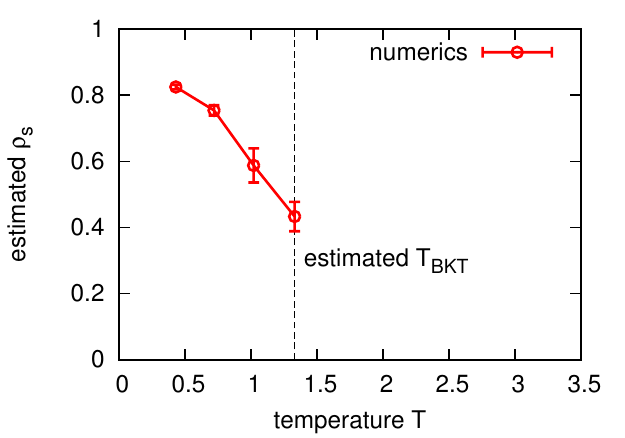}
\caption{
(Color online) Estimated superfluid fraction $ \rho_s $ with respect to different values of temperature.
The superlfuid fraction is evaluated by fitting the power-law exponents in Fig. \ref{correlationSize} and inverting equation (\ref{eq:nc-g1}).
Error bars indicates the measured standard deviation of the estimated superfluid fraction.
\label{superfluidTemperature}}
\end{figure}

We are now in the position to make estimations of $\rho_s$ at different temperatures by equating the measured slopes on the log-log 
plots of $g_1(r)$ with the value of $\alpha$ in the second equation (\ref{eq:nc-g1}), ie. $\alpha = T/( {4\pi \rho_s}) $. For this, we first plot on Fig. \ref{correlationSize} the log-log 
plots of $g_1(r)$ for different $T$ and $L$. 
We can see that the resulting curves are nearly independent of $L$ which confirm $L$-independence of $\rho_s$.
Next, measuring the slopes on these curves (with their relatives errors), we calculate $ \rho_s $ at different temperatures; the resulting dependence is shown in Fig. \ref{superfluidTemperature}. 


\subsection{Spectra
\label{subsec:spectra}} 

\begin{figure}
\includegraphics[width=0.8\linewidth]{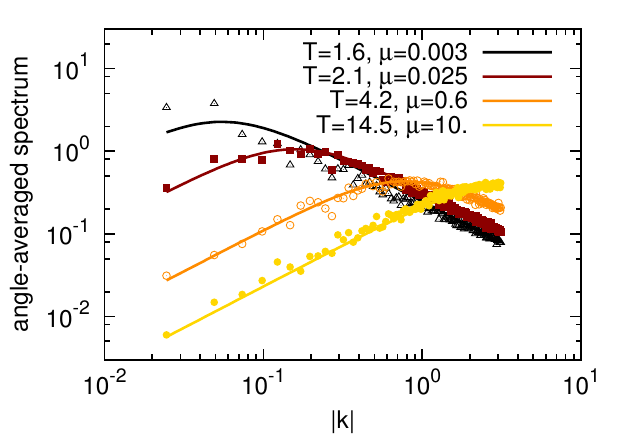} 
\caption{
(Color online) Angle-averaged steady state spectra (points) measured in different simulations having high free-particle energy densities and negligible condensate fraction.
The spectra are fitted with curves corresponding to the Rayleigh-Jeans distribution
(\ref{eq:rj}) integrated over angles, that is $ N_k= \int_0^{2\pi} n_k \, k \,  d\theta $.
Corresponding fitted values of temperature and chemical potential are shown in the labels.
Results are taken from simulations with $ L=256 $.
\label{fig:RJSpectra}}
\end{figure}

\begin{figure}
\includegraphics[width=0.8\linewidth]{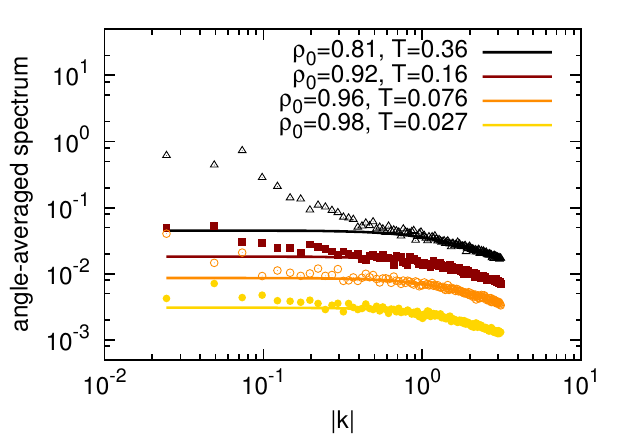} 
\caption{
(Color online) Angle-averaged steady state spectra (points) of Bogoliubov normal variables (\ref{bogolubov}) measured in different simulations having low free-particle energy densities and strong condensate fraction.
The spectra are fitted with curves corresponding to the Rayleigh-Jeans distribution (\ref{eq:RJ-bog}) integrated over angles, that is $ N_k= \int_0^{2\pi} n_k \, k \,  d\theta $.
Corresponding fitted values of temperature and condensate fraction are shown in the labels.
Results are taken from simulations with $ L=256 $.
\label{fig:BogoliubovSpectra}} 
\end{figure}

Now we will consider the steady state spectra  in simulations 
with $ L=256 $ and different values of energy.
Figure \ref{fig:RJSpectra} shows the spectra in simulations with high energies
where no condensation is observed together with the best fits obtained using the Rayleigh-Jeans  equilibrium distributions (\ref{eq:rj}). 
One can appreciate an excellent agreement between the data and the fits.
Moreover, relation between the fitted values $T$ and $\mu$ in Fig. \ref{fig:RJSpectra}
is in perfect agreement with theoretical formula
(\ref{eq:rho-of-T-mu}). 

Figure \ref{fig:BogoliubovSpectra} shows the spectra of the Bogoliubov quasi-particles in simulations with low energies
where a strong condensate {is present}, superimposing the best fits using the Rayleigh-Jeans equilibrium distributions
(\ref{eq:RJ-bog}). 
One can see a reasonable agreement between the data and the fits:
this agreement is better for lower temperatures and stronger condensate as the perturbations are better described by the Bogoliubov quasi-particles.
Also, these fits get better in the range of higher $k$'s, which contains most
of the systems energy, allowing a rather precise fit of $T$.

\begin{figure}
\includegraphics[width=0.8\linewidth]{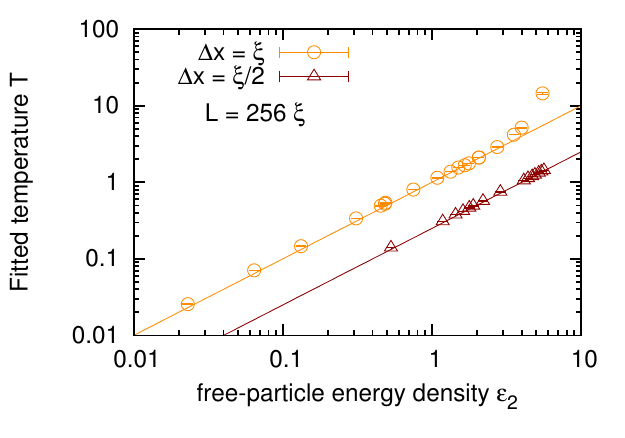} 
\caption{
(Color online) Temperatures (points) evaluated by fitting with the Rayleigh-Jeans distributions the final steady states with respect to the free-particle energy density $ \epsilon_2 $ (see Fig.s \ref{fig:RJSpectra} and \ref{fig:BogoliubovSpectra} for the corresponding values).
Two sets of simulations with same system size $ L=256 $ but different resolution ($ \Delta x=\xi $ and $ \Delta x = \xi/2 $) are shown.
The straight lines correspond to prediction (\ref{eq:epsilon_Tmu}) where the term containing the chemical potential $ \mu $ has been neglected.
Points out of the lines correspond to thermodynamic states where $ \mu \neq 0 $ thus having no condensate fraction.
Results are taken from simulations with $ L=256 $.
\label{fig:temperatureFromSpectra}}
\end{figure}

Figure \ref{fig:temperatureFromSpectra} plots the estimated temperature $T$, obtained from the fits
in Fig.s \ref{fig:RJSpectra} and \ref{fig:BogoliubovSpectra}, as a function of the free-particle 
energy density $\epsilon_2$.
Two different data sets are considered here: simulations where the resolution is 
$ \Delta x= \Delta y =\xi $ (as in all other results shown up to here) and simulations where the resolution
has been doubled, that is $ \Delta x'= \Delta y' =\xi/2 $ and consequently $ k_{\max}' = k_{\max} /2 $.
With that choice we are able to explore how the final steady states and their corresponding 
temperatures may depend on the ultraviolet cut-off $ k_{\max} $.    

Let us focus first on the former data set, that is where $ \Delta x= \Delta y =\xi $.
Remarkably, for both the low- and the high-temperature ranges this plot is 
in an excellent agreement with the 
expression (\ref{eq:epsilon_Tmu}) for $\epsilon_2$ 
in terms of $T$ and $\mu$, which was theoretically obtained for the high-energy case.
To understand why this expression works for the low-energy case, we first of all  
 notice that in the low-energy case most energy comes from the range
$k > k_a =k_{\max}/5.2$ (\ref{eq:aVortexCore}). Indeed, since the energy in the thermodynamic state is equipartitioned
over the 2D {\bf k}-space, the energy in the range $k > k_a $ represents about 97\% of the
total energy. But in this range $\omega_k \approx k^2$ and, according to formula (\ref{eq:ak-frre-partic}),
we have
$\langle |\hat a_{\bf k} |^2\rangle \approx \langle |\hat \psi_{\bf k} |^2\rangle.$
Thus, the RJ distribution in this range is the same as in for high-energy case with $\mu=0$, see (\ref{eq:rj}), and
therefore expression (\ref{eq:epsilon_Tmu}) is also the same (with $\mu=0$). 
Further, the second term on the RHS of expression (\ref{eq:epsilon_Tmu}), i.e. the one with $\mu$, is only
important for very high temperatures. In particular, all data points on  Fig. \ref{fig:temperatureFromSpectra}
except to the one corresponding to the highest $T$ fall onto the orange line
\begin{equation}
\label{TvsE}
T=\epsilon_2,
\end{equation}
which is nothing but expression (\ref{eq:epsilon_Tmu}) without the second term on the RHS and with $k_{\max}=\pi$.
This is a remarkably simple expression which has allowed us to identify $T$ and $\epsilon_2$ in all of our numerical experiments
(except for the highest $T$). Moreover, even for high $T$ expression (\ref{eq:epsilon_Tmu}) provides an excellent fit if one retains its second term
(with $\mu \ne 0$).

\begin{figure}
\includegraphics[width=0.8\linewidth]{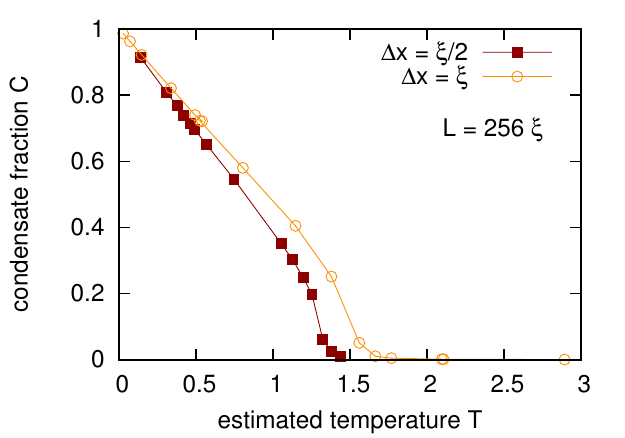} 
\caption{
(Color online) Measured condensate fraction with respect to the estimated temperatures fitted from spectra for systems having same size $ L=256 \xi $ but different resolution $ \Delta x=\xi $ and $ \Delta x=\xi/2 $.  
Results indicate that a better resolved system manifests a lower condensation temperature.
\label{fig:condensateTemperatureResolution}}
\end{figure}

For the runs with the doubled resolution, $ \Delta x= \Delta y =\xi/2 $, fitted temperatures
vs the free-particle energy densities are shown in Fig. \ref{fig:temperatureFromSpectra} with dark red points.
As the temperature is related to the average energy per mode, it is natural to expect four times lower $T$ for  the same free-particle energy densities  with respect to the lower resolution runs as the number of modes has now quadrupled.
This prediction given by the expression (\ref{eq:epsilon_Tmu}) divided by four is plotted with dark red line (neglecting again the chemical potential) in Fig. \ref{fig:temperatureFromSpectra} and it fits 
well the numerical results.
How does the resolution  affect the other results?
In Fig. \ref{fig:condensateTemperatureResolution} we plot  the condensate fraction with respect to the free-particle energy density in the two different resolution simulations.
Differences in the condensation temperatures are small but visible.
We interpret such dependence with the fact that with the increased resolution we are better resolving small scales  comparable to the vortex cores and the condensation process strongly depends on vortex interactions. 
We expect however that further increases in the resolution will lead to lower  changes converging to a cut-off independent limit as predicted in \cite{Holzmann30012007}.

\subsection{Vortices
\label{subsec:vortices}} 

Let us now look at the vortices, i.e. the zeroes of the field $\psi$ where the phase changes by $ \pm2\pi $.
It is well-known that NLS vortices behave similarly to hydrodynamic 
vortices if they are separated from each other by distances much
greater than the healing length $\xi$.
For example, two vortices of the same sign would rotate around
each other and two vortices of the opposite signs (a vortex dipole) would move 
along a straight line. 
However, when vortex separations are less
than $\xi$ they behave very different from their hydrodynamic counterparts.
A ``plasma" of  tightly placed ``ghost" vortices is simply a null set of a weakly nonlinear
dispersive wave field. There is no closed description of such a 
field in terms of the vortices only. In particular, the log-like expression for the
vortex energy used in the physical BKT argument reproduced in section \ref{sec:bkt})
would be invalid in this case.
Instead, a more fundamental physical quantity
in this case is the wave spectrum whose evolution satisfies the kinetic
equation (\ref{ke-nls}).

\begin{figure}
\includegraphics[scale=1.28]{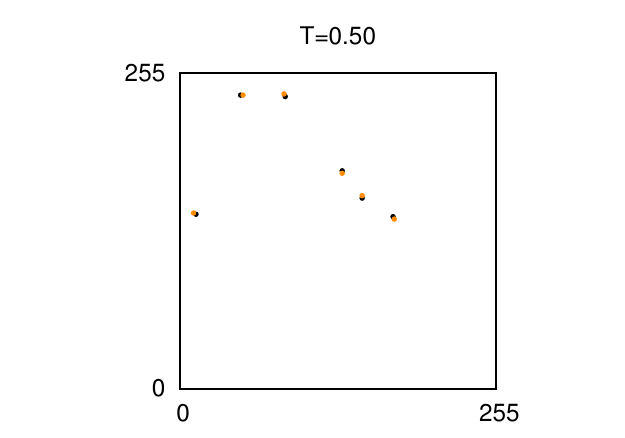}
\includegraphics[scale=1.28]{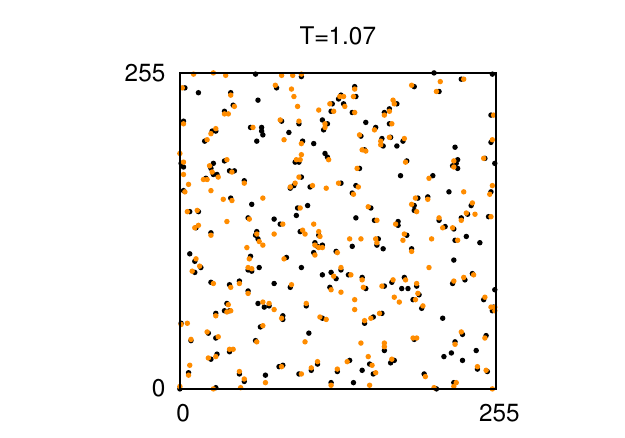}
\includegraphics[scale=1.28]{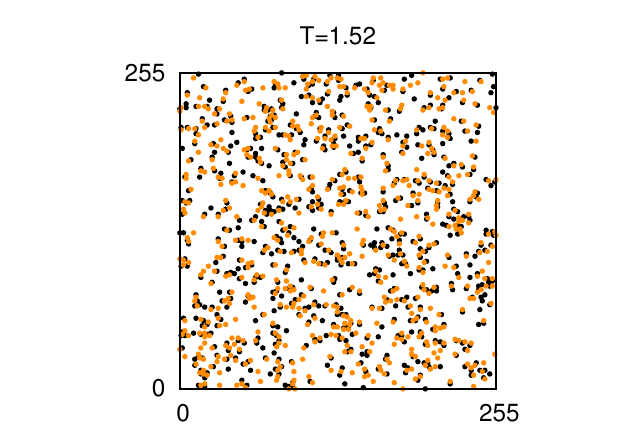}
\includegraphics[scale=1.28]{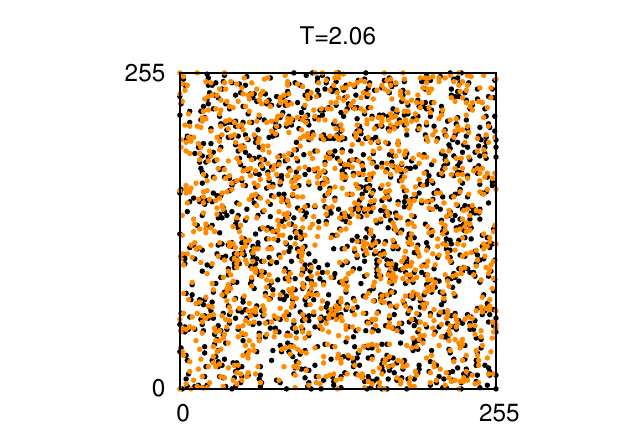}
\caption{
(Color online) Numerically detected quantum vortices in the final steady states for four simulations with $ L=256 $ and different temperatures.
Clockwise and anti-clockwise vortices are shown by dark yellow and black colours respectively.
Vortex number increases when increasing the temperature and clear hydrodynamic dipoles are only observed at very low temperatures, well below the estimated $ T_{BKT} $.
\label{fig:vortices}}
\end{figure}

\begin{figure}
\includegraphics[width=0.8\linewidth]{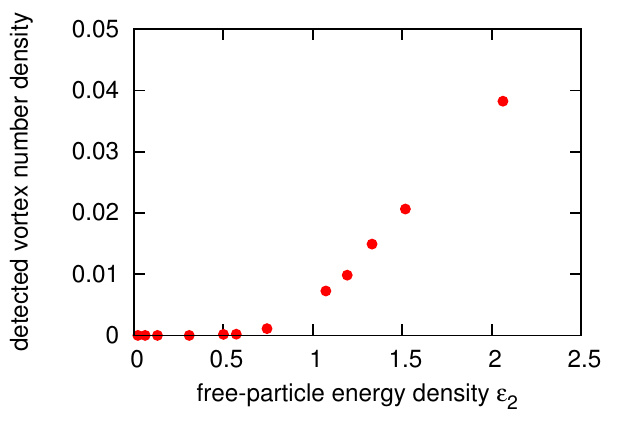}
\caption{
(Color online) Numerically measured detected number of vortices per density for different
values of the free-particle energy density.
Results are taken from simulations having side equal to $ L=256 $.
\label{fig:vorticesN} }
\end{figure}

Figure \ref{fig:vortices} shows 2D frames in the physical space with positive and negative vortices taken from simulations with $ L=256 $ and different $\epsilon_2$, that is different temperatures.
Figure \ref{fig:vorticesN} shows a plot of the number of vortices as a function
of $\epsilon_2$ in the same simulations.
As expected, we see that the number of vortices is decreased when the system is cooled,
and indeed we see formation of vortex dipoles at low $T$.
However, this transition is not sharp, and the dipoles are dominant only for temperatures which are significantly
lower than $T_{BKT}$, see the frame corresponding to $\epsilon_2 \approx T =0.5$ in
Figure \ref{fig:vortices}).  
For $T \lesssim T_{BKT} \approx 1.39$
the dipoles are in minority; for $\epsilon_2 \approx T =1.07$ we see,  apart from several dipoles,
several remaining isolated vortices as well as some
clusters of positive and negative vortices of larger sizes, with three or more vortices in them.
In fact, vortex clusters are seen also for $T \gtrsim T_{BKT}$, for instance for $\epsilon_2 \approx T =1.52$.
A numerical estimation of the vortex density has been previously done in
\cite{Nazarenko20061,nazon2007,PhysRevA.76.013613, PhysRevA.81.023623} but, as far as we know, 
such vortex clustering phenomenon of  has never been discussed in the context
of the BKT transition, and its possible dynamical role is yet to be understood.

In our simulation it is clear, however, that most vortices, in pairs, in clusters or in a vortex ``plasma", are not
separated by distances larger than $\xi$. 
Watching them in dynamics on a video reveals that these
vortices sporadically flicker in and out rather than getting engaged in a hydrodynamic type of 
motion (see movies in supplementary material). 
These are clearly ghost vortices representing short-wave fluctuations of a weakly nonlinear wave field, (waves for the high $T$ regime, Bogoliubov excitations in the low $ T $ regime), and their description in terms of the hydrodynamic Hamiltonian is invalid.
With these conclusions, it becomes even more puzzling why the physical argument based on 
the hydrodynamic Hamiltonian lead to an accurate BKT prediction for the power-law decay of correlations with the near-transition exponent $\alpha =1/4$.

\section{Conclusions \label{sec:conclusion}} 

In this paper, we studied statistical equilibrium states in the 2D NLS model which
is truncated in the Fourier space. No forcing or dissipation was introduced;
we let the nonlinearity playing the role of phase-space mixing and bringing the system to the 
thermodynamic equilibrium.
Our goal was to study the physical processes associated with
random interacting waves and vortices which accompany the Bose-Einstein
condensation and  the BKT transition and examine feasibility of the standard physical
assumptions put into the derivation of the power-law correlations and the BKT transition.

Because the NLS equation is also a universal model widely used in wave turbulence (WT),
we would like to establish closer links of the condensed matter theory of the above phenomena
with the WT approach and interpretations. 
Our choice of fully conservative formulation allowed us to look at the nonlinear dynamics of waves
and vortices in their purity and unobstructed by external stochasticity
associated with a thermal bath. 
In this respect our approach was different from
the one in \cite{PhysRevA.81.023623} where a model of  the incoherent field 
 was also added to the system.

Our numerical results support the view that the 2D NLS system in thermal
equilibrium represents a four-wave weak WT described by the kinetic equation
(\ref{ke-nls}) at  high temperatures
and a three-wave weak WT described by the kinetic equation
(\ref{kineq-R}) at  low temperatures. 
The assumption of the weaknessess of WT was confirmed using $(k,\omega)$-plots 
where it manifested itself as
narrow distributions around the linear-wave dispersion curves.
Importantly, not only the exponential decay of correlations at high $T$,
but also the power-law decay at low $T$ could be described within the WT approach.
For that WT has to be reintroduced at low $T$ using the hydrodynamic formulation,
which allows to use WT in cases where the phase experiences order one or greater changes.

At the intermediate temperatures,
i.e. the range in which the condensation and the BKT transition occur,
the system can be viewed as strong turbulent state in which the linear and the nonlinear
effects balance. In fact, it is precisely the balance of the linear and the nonlinear
terms in the NLS equation that appears to give the correct criterion for the
condensation and the BKT transition, and the respective transition temperatures
can be found from this condition. 
To be precise, we found that for the same system size, provided is large enough (for example $ L=256 $, where the length unit is in term of the healing length),  the condensation
temperature $T_c$ is, within the error bar, the same as $T_{BKT} \approx 1.39$,
and that this temperature corresponds to the ratio of the interaction and the free-particle
energies $ \sigma \approx 0.5$.
For much smaller finite size $L$ we find that $ T_c > T_{BKT} $ similarly to what was observed before in numerical 
experiments of trapped Bose gases \cite{PhysRevLett.100.190402, PhysRevA.79.033626}.
We recall also that when the system is sufficiently small then a macroscopic occupation of the lowest energy mode occurs even in the limit of non interacting gas \cite{pitaevskii2003bose, PhysRevA.81.023623}.

To fully understand the role of WT processes in the BKT transition, it is useful to consider a non-equilibrium setup where WT description works at its best. Namely, it is useful to consider the transient non-equilibrium kinetics before relaxation to the thermal equilibrium. For that let us imagine a system with initial excitations in some limited range of intermediate wavenumbers. The wave turbulence (similarly to 2D classical turbulence) predicts two cascades: an energy cascade toward higher momenta, which can be qualitatively associated with evaporative cooling mechanism, and a particle cascade toward lower momenta - a precursor of the condensation. Both of these cascades are well described by the 4-wave kinetic equation of wave turbulence \cite{DNPZ-bec,Nazarenko20061,nazon2007}. During the particle cascade, for sufficiently low temperature, the weakness of interaction assumption breaks down and the system enters into a strongly nonlinear stage characterized by interacting vortices, which tend to annihilate on average \cite{Nazarenko20061,nazon2007}. After that, for sufficiently low temperature, the system enters into another wave turbulence regime, now 3-wave type, characterised by interacting Bogoliubov phonons \cite{DNPZ-bec,Zakharov2005203,Nazarenko20061}. It is precisely in between when the temperature is sufficiently low for the 4-wave turbulence to break down but not low enough to enter into the the 3-wave turbulence regime in the equilibrium state that the BKT transition occurs. In the Gross-Pitaevskii equation this corresponds to the equality of the linear and the nonlinear terms, which was confirmed with high accuracy in our numerical results.

For the equilibrium state, we have found some striking confirmations of the predictions
of the power-law correlation decay and of its exponent at the BKT transition temperature,
$\alpha =1/4$, as well as a reasonable agreement with the prediction
of the relation between the superfluid density and the transition temperature,
$T_{BKT} = \pi \rho_s $.
Moreover, we observed that the superfluid density $ \rho_s $ does not depend on the system size $ L $ 
and, 
as observed for trapped gases in \cite{PhysRevLett.100.190402},
the condensate fraction $ \rho_0 $ scales like a power-law with respect to $ L $ above the BKT transition.
We stress however that our definition  of the superfluid density $ \rho_s $ via postulating the model Hamiltonian in eq. (\ref{eq:hydro-H-rhos}) is different from a common definition of $ \rho_s $  by considering superfluid stiffness or reduced moment of inertia \cite{PhysRevA.81.023623}.
Our definition leads to the prediction of 
 the power-law behaviour of the first-order correlation function, which we have validated numerically and thereby  indirectly found the values of  $ \rho_s $ by fitting such power laws.
The question how well the two definitions of   $ \rho_s $  correspond with each other has not been considered in the present paper and it deserves further analysis. 

At the same time, we found inconsistency of the basic physical assumptions used for deriving the 
power-law correlations near the BKT transition temperatures (i.e. postulating 
the superfluid density and thereby redefining the hydrodynamic 
and Bogoliubov approach) as well as the BKT threshold (i.e. using the picture of hydrodynamic
vortices with log-like expressions for their energies). First, the frequency shifts for $k \to 0$
obtained from the
$(k,\omega)$-plots contradict the naive view that for intermediate temperatures
the system can be described by Bogoliubov approach with some effective superfluid density.
Second, at the BKT transition temperature (as well as at temperatures close to it) the NLS vortices
have properties very different from their hydrodynamic counterparts and there is no closed description
of the system in terms of these vortices only. The vortices are not separated by distances $>\xi$ sufficient
for them to become self-sustained and hydrodynamic-like. 
Also, the physical picture that at crossing  $T_{BKT}$ the system changes from being a plasma
of individual vortices to a gas of bound vortex dipoles appears to be oversimplified.
The transition appears to be rather gradual and in a wide range of temperatures around $T_{BKT}$
one observes vortex clusters. The mean size of such vortex clusters gradually decreases at $T$
is decreased, and only of temperatures significantly lower than $T_{BKT}$ simple vortex dipoles start to dominate.
Moreover, quite small further decrease of $T$ leads to disappearance of vortices altogether.

It is often said that more rigorous and systematic description of the BKT transition exists within the renormalization 
group approach. However, as we have sees in the present paper, more work remains to be done for
understanding the underlying physical properties in terms of interacting waves, vortices and or/and other
nonlinear hydrodynamic structures.

\appendix

\section{Assumptions to derive 2D NLS for a Bose gas \label{app:NLS-BEC}}

In the following we shall explain when the NLS equation could be used as a model for a Bose gas at finite low temperatures and obtain the rescaling coefficients to link the real physical quantities to the non-dimensional ones used in our work.
To make a clear distinction between them we will indicate non-dimensional quantities using the superscript $ (\cdot)^{(nd)} $.
Note that in the main text this superscript has been omitted not to overweight the notation.

To study weakly interacting bosons it is useful to work in the framework of second quantisation and introduce the many body Hamiltonian operator \cite{pitaevskii2003bose}
\begin{equation}
\hat{H} = \int \frac{\hbar^2}{2 m} \nabla\hat{\Psi}(\mathbf{r})  \nabla\hat{\Psi}^\dagger(\mathbf{r}) \, d\mathbf{r} \, + \int \hat{\Psi}^\dagger (\mathbf{r}) \hat{\Psi}^\dagger (\mathbf{r'}) V(\mathbf{r}-\mathbf{r'}) \hat{\Psi} (\mathbf{r}) \hat{\Psi} (\mathbf{r'}) \, d\mathbf{r} d\mathbf{r'} \, .
\end{equation}
Here $ \hbar $ is the Planck's constant, $ m $ is the boson mass, $ \hat{\Psi}(\mathbf{r}) $ and $ \hat{\Psi}^\dagger(\mathbf{r}) $ are the destruction and creation quantum operators respectively, and $ V(\mathbf{r}-\mathbf{r'}) $ represents the two-particle interaction potential.
If for simplicity a $ \delta $-form potential $ V(\mathbf{r}-\mathbf{r'})=V_0 \, \delta(\mathbf{r}-\mathbf{r'}) $ is considered, the temporal evolution of the destruction operator results simply in
\begin{equation}
i \hbar \, \frac{\partial \hat{\Psi}(\mathbf{r})}{\partial t} = - \frac{\hbar^2}{2 m} \nabla^2 \hat{\Psi}(\mathbf{r}) + V_0 \hat{\Psi}^\dagger(\mathbf{r}) \hat{\Psi}(\mathbf{r}) \hat{\Psi}(\mathbf{r}) \, .
\label{eq:opGPE}
\end{equation}
Following \cite{PhysRevLett.87.160402}, the field operator $ \hat{\Psi} $ can be expressed in the momentum space $ \mathbf{k} $ using a projector operator $ \hat{\mathcal{P}} $. The decomposition follows
\begin{equation}
\hat{\mathcal{P}} \hat \Psi(\mathbf{r}) = \sum_{\mathbf{k} \, \in \, C} \hat{a}_{\mathbf{k}} \, \phi_{\mathbf{k}}(\mathbf{r}) \, ,
\end{equation}
where the region $ C $ is chosen to satisfy the requirement that the expected occupation number $ N_\mathbf{k} = \langle \hat{a}_\mathbf{k}^\dagger \hat{a}_\mathbf{k} \rangle \gg 1 $ and the set $ \left\{ \phi_{\mathbf{k}} \right\} $ defines a basis in which the field operator is approximately diagonal at the boundary of $ C $.
For these modes quantum fluctuations can be ignored and thus the operator $ \hat{a}_{\mathbf{k}} $ can be replaced by a complex number $ c_{\mathbf{k}} $ so that
\begin{equation}
\hat{\mathcal{P}} \hat \Psi(\mathbf{r}) \longrightarrow \psi(\mathbf{r}) = \sum_{\mathbf{k} \, \in \, C} c_{\mathbf{k}} \, \phi_{\mathbf{k}}(\mathbf{r}) \, .
\end{equation}
In the hypothesis that the contributions of $ (\hat{\mathds{1} }- \hat{\mathcal{P}}) \hat{\Psi}(\mathbf{r}) $ are negligible and that the region $ C $ is large enough that the fluctuations of particle and energy are small enough in the grand canonical ensemble, equation (\ref{eq:opGPE}) simplifies to
\begin{equation}
i \hbar \, \frac{\partial \psi}{\partial t}(\mathbf{r}, t) = - \frac{\hbar^2}{2 m} \nabla^2 \psi(\mathbf{r}, t) + V_0 |\psi(\mathbf{r}, t)|^2 \psi(\mathbf{r}, t) \, .
\end{equation}
This is nothing but the three-dimensional nonlinear Schr{\"o}dinger equation well-known as Gross-Pitaevskii equation in the BEC community.
We underline however that usually the Gross-Pitaevskii equation describes the dynamics of the order parameter $ \psi $ in the limit of the absolute temperature $ T=0 $, that is when all bosons already occupies the single particle lowest energy level state. 
On the contrary, the derivation presented above let us extend this mean-field model to finite (but still low) temperatures, provided that all projected modes are highly occupied.
In this limit the nonlinear term in the equation is sufficient to drive the system towards the thermodynamic equilibrium, thus no external thermal bath is needed.

A three-dimensional Bose gas can be squeezed into a (quasi-)two-dimensional one if only one degree of freedom is accessible in the third dimension. 
This can be experimentally realised using a tight confining harmonic potential $ V_{ext}(z)=1/2 \, m \omega_z^2 z^2 $ is applied for example in the $ z $-direction.
Recalling that the characteristic oscillator length is $ a_z=\sqrt{\hbar/m\omega_z} $ and the interacting potential between bosons, $ V_0 = 4\pi \hbar a_s/m $, depends on the particle scattering length $ a_s $, the system is then modelled by the two-dimensional nonlinear Schr{\"o}dinger equation
\begin{equation}
i \hbar \, \frac{\partial \psi_{2D}}{\partial t}(\mathbf{x}, t) = - \frac{\hbar^2}{2 m} \nabla_{2D}^2 \psi_{2D}(\mathbf{x}, t) + \frac{\sqrt{8\pi} \hbar^2 a_s}{m a_z} |\psi_{2D}(\mathbf{x}, t)|^2 \psi_{2D}(\mathbf{x}, t) - \mu_{V} \psi_{2D}(\mathbf{x}, t) \, , 
\label{eq:2DNLS-dim}
\end{equation}
with $ \mathbf{x} \in \mathbb{R}^2 $ and $ \nabla_{2D} $ being the two-dimensional gradient. 
Here $ \psi_{2D} = \sqrt{a_z} \, \psi $ and $ \mu_{V} = \hbar \omega_z/2 $ is the chemical potential due to the $ z $-axis confinement.
Its non-dimensional version can be easily obtained by rescaling the following quantities
\begin{equation}
\left\{
\begin{split}
& \mathbf{x} = \delta \, \mathbf{x}^{(nd)} \\
& t = \frac{\hbar}{\epsilon_\delta} \, t^{(nd)} \\
& \psi_{2D} = \sqrt{n_2} \, e^{i \frac{\omega_z}{2} t} \, \psi_{2D}^{(nd)} \quad \Longrightarrow \quad \psi = \sqrt{\frac{n_2}{a_z}} \, e^{i \frac{\omega_z}{2} t} \, \psi_{2D}^{(nd)} \\
\end{split}
\right. \, .
\end{equation} 
Here we set $ n_2 = N/(L_x L_y) $ the two-dimensional N bosons number density of a system having size $ L_x \times L_y \times a_z $, $ \delta = \sqrt{a_z /(8\pi a_s n_2)} $ the characteristic length at which linear and nonlinear terms in the NLS equation (\ref{eq:2DNLS-dim}) are comparable, and $ \epsilon_\delta = \hbar^2 / (2 m \delta^2) $ its characteristic quantum-mechanical energy.

Finally we recall that a system of non-interacting (or weakly-interacting) scalar bosons in equilibrium with a thermal bath naturally follows the Bose-Einstein statistics. 
The expected occupation number of a quantum state having a characteristic wavenumber $ \mathbf{k} $ results in
\begin{equation}
n_{BE}(\mathbf{k})=\frac{1}{e^{\frac{\epsilon(\mathbf{k})+\mu}{k_B T}}-1} \, ,
\end{equation}
where $ \epsilon(\mathbf{k}) $ is the single particle energy, $ k_B $ is the Boltzmann's constant, $ \mu $ is the chemical potential and $ T $ is the absolute temperature.
In the limit of very small wavenumbers, i.e. for $ \epsilon(\mathbf{k})+\mu \ll k_B T $, it is straightforward to prove that the Bose-Einstein statistics reduces to the Rayleigh-Jeans statistics
\begin{equation}
n_{RJ}(\mathbf{k})= \frac{k_B T}{\epsilon(\mathbf{k})+\mu} \, .
\label{eq:dimRJ}
\end{equation}
Using the previous rescaling factors it is easy to prove that
\begin{equation}
n_{RJ}(\mathbf{k}) = n_2 \delta^2 \, n_{RJ}^{(nd)}(\mathbf{k}^{(nd)})
\end{equation} as the total boson number N is 
\begin{equation}
\begin{split}
\int \frac{k_B T}{\epsilon(\mathbf{k}) + \mu_{2D} +\mu_V} \, dk_x dk_y & = n_2 \int \frac{T^{(nd)}}{\epsilon^{(nd)}(\mathbf{k}^{(nd)}) + \mu^{(nd)}_{2D}} dk^{(nd)}_x dk^{(nd)}_y \\ 
\int \frac{1}{\delta^2} \frac{k_B T}{\epsilon(\mathbf{k}) + \mu_{2D} + \mu_V} \, dk^{(nd)}_x dk^{(nd)}_y & = \int n_2 \frac{T^{(nd)}}{\epsilon^{(nd)}(\mathbf{k}^{(nd)}) + \mu^{(nd)}_{2D}} dk^{(nd)}_x dk^{(nd)}_y \, . \\ 
\end{split}
\end{equation}
Thus, if we rescale energy and chemical potential as $ \epsilon (\mathbf{k}) + \mu_V = \epsilon_{\delta} \, \epsilon^{(nd)}(\mathbf{k}^{(nd)}) $ and $ \mu_{2D}= \epsilon_{\delta} \, \mu_{2D}^{(nd)} $, the temperature rescales as
\begin{equation}
T = \frac{\delta^2 n_2 \epsilon_{\delta}}{k_B} \, T^{(nd)} = \frac{N \hbar^2}{2 m L_x L_y k_B} \, T^{(nd)} \,  .
\end{equation}

A final important remark should be made: in our model we only consider the coherent region c-field of the Bose gas, without taking into account the incoherent field operator.
The inclusion of the incoherent field operator changes the evaluation of the total number of particles or the total energy of the Bose gas as shown in \cite{PhysRevA.81.023623}.

\begin{acknowledgments}
Authors wish to express their gratitude to C.F.~Barenghi, D.~Gangardt, N.~Proukakis, H.~Salman, and Y.~Sergeev for interesting discussions and suggestions.
The CINECA Award (No. HP10BQW4X9) 2011 is also aknowledge for the availability of high-performance computing resources and support.
Part of the numerical results presented in this paper were also carried out on the High Performance Computing Cluster supported by the Research and Specialist Computing Support service at the University of East Anglia.
S.~N. acknowledges support from the government of Russian Federation via grant
No. 12.740.11.1430 for supporting research of teams working under supervision of invited scientists.
For the second part of this project S.~N. was funded by a Chaire Senior PALM ``TurbOndes''.

\end{acknowledgments}


\bibliography{references}

\begin{thebibliography}{57}%
\makeatletter
\providecommand \@ifxundefined [1]{%
 \@ifx{#1\undefined}
}%
\providecommand \@ifnum [1]{%
 \ifnum #1\expandafter \@firstoftwo
 \else \expandafter \@secondoftwo
 \fi
}%
\providecommand \@ifx [1]{%
 \ifx #1\expandafter \@firstoftwo
 \else \expandafter \@secondoftwo
 \fi
}%
\providecommand \natexlab [1]{#1}%
\providecommand \enquote  [1]{``#1''}%
\providecommand \bibnamefont  [1]{#1}%
\providecommand \bibfnamefont [1]{#1}%
\providecommand \citenamefont [1]{#1}%
\providecommand \href@noop [0]{\@secondoftwo}%
\providecommand \href [0]{\begingroup \@sanitize@url \@href}%
\providecommand \@href[1]{\@@startlink{#1}\@@href}%
\providecommand \@@href[1]{\endgroup#1\@@endlink}%
\providecommand \@sanitize@url [0]{\catcode `\\12\catcode `\$12\catcode
  `\&12\catcode `\#12\catcode `\^12\catcode `\_12\catcode `\%12\relax}%
\providecommand \@@startlink[1]{}%
\providecommand \@@endlink[0]{}%
\providecommand \url  [0]{\begingroup\@sanitize@url \@url }%
\providecommand \@url [1]{\endgroup\@href {#1}{\urlprefix }}%
\providecommand \urlprefix  [0]{URL }%
\providecommand \Eprint [0]{\href }%
\providecommand \doibase [0]{http://dx.doi.org/}%
\providecommand \selectlanguage [0]{\@gobble}%
\providecommand \bibinfo  [0]{\@secondoftwo}%
\providecommand \bibfield  [0]{\@secondoftwo}%
\providecommand \translation [1]{[#1]}%
\providecommand \BibitemOpen [0]{}%
\providecommand \bibitemStop [0]{}%
\providecommand \bibitemNoStop [0]{.\EOS\space}%
\providecommand \EOS [0]{\spacefactor3000\relax}%
\providecommand \BibitemShut  [1]{\csname bibitem#1\endcsname}%
\let\auto@bib@innerbib\@empty
\bibitem [{\citenamefont {Pitaevskii}\ and\ \citenamefont
  {Stringari}(2003)}]{pitaevskii2003bose}%
  \BibitemOpen
  \bibfield  {author} {\bibinfo {author} {\bibfnamefont {L.}~\bibnamefont
  {Pitaevskii}}\ and\ \bibinfo {author} {\bibfnamefont {S.}~\bibnamefont
  {Stringari}},\ }\href {http://books.google.co.uk/books?id=rIobbOxC4j4C}
  {\emph {\bibinfo {title} {Bose-Einstein Condensation}}},\ International
  Series of Monographs on Physics\ (\bibinfo  {publisher} {Clarendon Press},\
  \bibinfo {year} {2003})\BibitemShut {NoStop}%
\bibitem [{\citenamefont {Mermin}\ and\ \citenamefont
  {Wagner}(1966)}]{MerminWagner}%
  \BibitemOpen
  \bibfield  {author} {\bibinfo {author} {\bibfnamefont {N.~D.}\ \bibnamefont
  {Mermin}}\ and\ \bibinfo {author} {\bibfnamefont {H.}~\bibnamefont
  {Wagner}},\ }\href {\doibase 10.1103/PhysRevLett.17.1133} {\bibfield
  {journal} {\bibinfo  {journal} {Physical Review Letters}\ }\textbf {\bibinfo
  {volume} {17}},\ \bibinfo {pages} {1133} (\bibinfo {year}
  {1966})}\BibitemShut {NoStop}%
\bibitem [{\citenamefont {Bogoliubov}\ and\ \citenamefont
  {Bogoliubov~(Jr.)}(2009)}]{Bogoliubov2009}%
  \BibitemOpen
  \bibfield  {author} {\bibinfo {author} {\bibfnamefont {N.~N.}\ \bibnamefont
  {Bogoliubov}}\ and\ \bibinfo {author} {\bibfnamefont {N.~N.}\ \bibnamefont
  {Bogoliubov~(Jr.)}},\ }\href@noop {} {\emph {\bibinfo {title} {Introduction
  to Quantum Statistical Mechanics}}}\ (\bibinfo  {publisher} {2nd ed., World
  Scientific, Singapore},\ \bibinfo {year} {2009})\BibitemShut {NoStop}%
\bibitem [{\citenamefont {Berezinskii}(1971)}]{Berezinskii}%
  \BibitemOpen
  \bibfield  {author} {\bibinfo {author} {\bibfnamefont {V.}~\bibnamefont
  {Berezinskii}},\ }\href@noop {} {\bibfield  {journal} {\bibinfo  {journal}
  {Sov. Phys. JETP}\ }\textbf {\bibinfo {volume} {32}},\ \bibinfo {pages} {493}
  (\bibinfo {year} {1971})}\BibitemShut {NoStop}%
\bibitem [{\citenamefont {{Kosterlitz}}\ and\ \citenamefont
  {{Thouless}}(1973)}]{KT}%
  \BibitemOpen
  \bibfield  {author} {\bibinfo {author} {\bibfnamefont {J.~M.}\ \bibnamefont
  {{Kosterlitz}}}\ and\ \bibinfo {author} {\bibfnamefont {D.~J.}\ \bibnamefont
  {{Thouless}}},\ }\href {\doibase 10.1088/0022-3719/6/7/010} {\bibfield
  {journal} {\bibinfo  {journal} {Journal of Physics C Solid State Physics}\
  }\textbf {\bibinfo {volume} {6}},\ \bibinfo {pages} {1181} (\bibinfo {year}
  {1973})}\BibitemShut {NoStop}%
\bibitem [{\citenamefont {Kosterlitz}(1974)}]{kosterlitz1974critical}%
  \BibitemOpen
  \bibfield  {author} {\bibinfo {author} {\bibfnamefont {J.}~\bibnamefont
  {Kosterlitz}},\ }\href@noop {} {\bibfield  {journal} {\bibinfo  {journal}
  {Journal of Physics C: Solid State Physics}\ }\textbf {\bibinfo {volume}
  {7}},\ \bibinfo {pages} {1046} (\bibinfo {year} {1974})}\BibitemShut
  {NoStop}%
\bibitem [{\citenamefont {Stock}\ \emph {et~al.}(2005)\citenamefont {Stock},
  \citenamefont {Hadzibabic}, \citenamefont {Battelier}, \citenamefont
  {Cheneau},\ and\ \citenamefont {Dalibard}}]{PhysRevLett.95.190403}%
  \BibitemOpen
  \bibfield  {author} {\bibinfo {author} {\bibfnamefont {S.}~\bibnamefont
  {Stock}}, \bibinfo {author} {\bibfnamefont {Z.}~\bibnamefont {Hadzibabic}},
  \bibinfo {author} {\bibfnamefont {B.}~\bibnamefont {Battelier}}, \bibinfo
  {author} {\bibfnamefont {M.}~\bibnamefont {Cheneau}}, \ and\ \bibinfo
  {author} {\bibfnamefont {J.}~\bibnamefont {Dalibard}},\ }\href {\doibase
  10.1103/PhysRevLett.95.190403} {\bibfield  {journal} {\bibinfo  {journal}
  {Phys. Rev. Lett.}\ }\textbf {\bibinfo {volume} {95}},\ \bibinfo {pages}
  {190403} (\bibinfo {year} {2005})}\BibitemShut {NoStop}%
\bibitem [{\citenamefont {Hadzibabic}\ \emph {et~al.}(2006)\citenamefont
  {Hadzibabic}, \citenamefont {Kr{\"u}ger}, \citenamefont {Cheneau},
  \citenamefont {Battelier},\ and\ \citenamefont
  {Dalibard}}]{hadzibabic2006berezinskii}%
  \BibitemOpen
  \bibfield  {author} {\bibinfo {author} {\bibfnamefont {Z.}~\bibnamefont
  {Hadzibabic}}, \bibinfo {author} {\bibfnamefont {P.}~\bibnamefont
  {Kr{\"u}ger}}, \bibinfo {author} {\bibfnamefont {M.}~\bibnamefont {Cheneau}},
  \bibinfo {author} {\bibfnamefont {B.}~\bibnamefont {Battelier}}, \ and\
  \bibinfo {author} {\bibfnamefont {J.}~\bibnamefont {Dalibard}},\ }\href@noop
  {} {\bibfield  {journal} {\bibinfo  {journal} {Nature}\ }\textbf {\bibinfo
  {volume} {441}},\ \bibinfo {pages} {1118} (\bibinfo {year}
  {2006})}\BibitemShut {NoStop}%
\bibitem [{\citenamefont {Kr\"uger}\ \emph {et~al.}(2007)\citenamefont
  {Kr\"uger}, \citenamefont {Hadzibabic},\ and\ \citenamefont
  {Dalibard}}]{PhysRevLett.99.040402}%
  \BibitemOpen
  \bibfield  {author} {\bibinfo {author} {\bibfnamefont {P.}~\bibnamefont
  {Kr\"uger}}, \bibinfo {author} {\bibfnamefont {Z.}~\bibnamefont
  {Hadzibabic}}, \ and\ \bibinfo {author} {\bibfnamefont {J.}~\bibnamefont
  {Dalibard}},\ }\href {\doibase 10.1103/PhysRevLett.99.040402} {\bibfield
  {journal} {\bibinfo  {journal} {Phys. Rev. Lett.}\ }\textbf {\bibinfo
  {volume} {99}},\ \bibinfo {pages} {040402} (\bibinfo {year}
  {2007})}\BibitemShut {NoStop}%
\bibitem [{\citenamefont {Schweikhard}\ \emph {et~al.}(2007)\citenamefont
  {Schweikhard}, \citenamefont {Tung},\ and\ \citenamefont
  {Cornell}}]{PhysRevLett.99.030401}%
  \BibitemOpen
  \bibfield  {author} {\bibinfo {author} {\bibfnamefont {V.}~\bibnamefont
  {Schweikhard}}, \bibinfo {author} {\bibfnamefont {S.}~\bibnamefont {Tung}}, \
  and\ \bibinfo {author} {\bibfnamefont {E.~A.}\ \bibnamefont {Cornell}},\
  }\href {\doibase 10.1103/PhysRevLett.99.030401} {\bibfield  {journal}
  {\bibinfo  {journal} {Phys. Rev. Lett.}\ }\textbf {\bibinfo {volume} {99}},\
  \bibinfo {pages} {030401} (\bibinfo {year} {2007})}\BibitemShut {NoStop}%
\bibitem [{\citenamefont {Clad\'e}\ \emph {et~al.}(2009)\citenamefont
  {Clad\'e}, \citenamefont {Ryu}, \citenamefont {Ramanathan}, \citenamefont
  {Helmerson},\ and\ \citenamefont {Phillips}}]{PhysRevLett.102.170401}%
  \BibitemOpen
  \bibfield  {author} {\bibinfo {author} {\bibfnamefont {P.}~\bibnamefont
  {Clad\'e}}, \bibinfo {author} {\bibfnamefont {C.}~\bibnamefont {Ryu}},
  \bibinfo {author} {\bibfnamefont {A.}~\bibnamefont {Ramanathan}}, \bibinfo
  {author} {\bibfnamefont {K.}~\bibnamefont {Helmerson}}, \ and\ \bibinfo
  {author} {\bibfnamefont {W.~D.}\ \bibnamefont {Phillips}},\ }\href {\doibase
  10.1103/PhysRevLett.102.170401} {\bibfield  {journal} {\bibinfo  {journal}
  {Phys. Rev. Lett.}\ }\textbf {\bibinfo {volume} {102}},\ \bibinfo {pages}
  {170401} (\bibinfo {year} {2009})}\BibitemShut {NoStop}%
\bibitem [{\citenamefont {Hung}\ \emph {et~al.}(2011)\citenamefont {Hung},
  \citenamefont {Zhang}, \citenamefont {Gemelke},\ and\ \citenamefont
  {Chin}}]{hung2011observation}%
  \BibitemOpen
  \bibfield  {author} {\bibinfo {author} {\bibfnamefont {C.-L.}\ \bibnamefont
  {Hung}}, \bibinfo {author} {\bibfnamefont {X.}~\bibnamefont {Zhang}},
  \bibinfo {author} {\bibfnamefont {N.}~\bibnamefont {Gemelke}}, \ and\
  \bibinfo {author} {\bibfnamefont {C.}~\bibnamefont {Chin}},\ }\href@noop {}
  {\bibfield  {journal} {\bibinfo  {journal} {Nature}\ }\textbf {\bibinfo
  {volume} {470}},\ \bibinfo {pages} {236} (\bibinfo {year}
  {2011})}\BibitemShut {NoStop}%
\bibitem [{\citenamefont {Ha}\ \emph {et~al.}(2013)\citenamefont {Ha},
  \citenamefont {Hung}, \citenamefont {Zhang}, \citenamefont {Eismann},
  \citenamefont {Tung},\ and\ \citenamefont {Chin}}]{PhysRevLett.110.145302}%
  \BibitemOpen
  \bibfield  {author} {\bibinfo {author} {\bibfnamefont {L.-C.}\ \bibnamefont
  {Ha}}, \bibinfo {author} {\bibfnamefont {C.-L.}\ \bibnamefont {Hung}},
  \bibinfo {author} {\bibfnamefont {X.}~\bibnamefont {Zhang}}, \bibinfo
  {author} {\bibfnamefont {U.}~\bibnamefont {Eismann}}, \bibinfo {author}
  {\bibfnamefont {S.-K.}\ \bibnamefont {Tung}}, \ and\ \bibinfo {author}
  {\bibfnamefont {C.}~\bibnamefont {Chin}},\ }\href {\doibase
  10.1103/PhysRevLett.110.145302} {\bibfield  {journal} {\bibinfo  {journal}
  {Phys. Rev. Lett.}\ }\textbf {\bibinfo {volume} {110}},\ \bibinfo {pages}
  {145302} (\bibinfo {year} {2013})}\BibitemShut {NoStop}%
\bibitem [{\citenamefont {Bagnato}\ and\ \citenamefont
  {Kleppner}(1991)}]{PhysRevA.44.7439}%
  \BibitemOpen
  \bibfield  {author} {\bibinfo {author} {\bibfnamefont {V.}~\bibnamefont
  {Bagnato}}\ and\ \bibinfo {author} {\bibfnamefont {D.}~\bibnamefont
  {Kleppner}},\ }\href {\doibase 10.1103/PhysRevA.44.7439} {\bibfield
  {journal} {\bibinfo  {journal} {Phys. Rev. A}\ }\textbf {\bibinfo {volume}
  {44}},\ \bibinfo {pages} {7439} (\bibinfo {year} {1991})}\BibitemShut
  {NoStop}%
\bibitem [{\citenamefont {Petrov}\ \emph {et~al.}(2000)\citenamefont {Petrov},
  \citenamefont {Holzmann},\ and\ \citenamefont
  {Shlyapnikov}}]{PhysRevLett.84.2551}%
  \BibitemOpen
  \bibfield  {author} {\bibinfo {author} {\bibfnamefont {D.~S.}\ \bibnamefont
  {Petrov}}, \bibinfo {author} {\bibfnamefont {M.}~\bibnamefont {Holzmann}}, \
  and\ \bibinfo {author} {\bibfnamefont {G.~V.}\ \bibnamefont {Shlyapnikov}},\
  }\href {\doibase 10.1103/PhysRevLett.84.2551} {\bibfield  {journal} {\bibinfo
   {journal} {Phys. Rev. Lett.}\ }\textbf {\bibinfo {volume} {84}},\ \bibinfo
  {pages} {2551} (\bibinfo {year} {2000})}\BibitemShut {NoStop}%
\bibitem [{\citenamefont {Prokof'ev}\ \emph {et~al.}(2001)\citenamefont
  {Prokof'ev}, \citenamefont {Ruebenacker},\ and\ \citenamefont
  {Svistunov}}]{Prokof2001}%
  \BibitemOpen
  \bibfield  {author} {\bibinfo {author} {\bibfnamefont {N.}~\bibnamefont
  {Prokof'ev}}, \bibinfo {author} {\bibfnamefont {O.}~\bibnamefont
  {Ruebenacker}}, \ and\ \bibinfo {author} {\bibfnamefont {B.}~\bibnamefont
  {Svistunov}},\ }\href@noop {} {\bibfield  {journal} {\bibinfo  {journal}
  {Phys. Rev. Lett.}\ }\textbf {\bibinfo {volume} {87}},\ \bibinfo {pages}
  {270402} (\bibinfo {year} {2001})}\BibitemShut {NoStop}%
\bibitem [{\citenamefont {Giorgetti}\ \emph {et~al.}(2007)\citenamefont
  {Giorgetti}, \citenamefont {Carusotto},\ and\ \citenamefont
  {Castin}}]{PhysRevA.76.013613}%
  \BibitemOpen
  \bibfield  {author} {\bibinfo {author} {\bibfnamefont {L.}~\bibnamefont
  {Giorgetti}}, \bibinfo {author} {\bibfnamefont {I.}~\bibnamefont
  {Carusotto}}, \ and\ \bibinfo {author} {\bibfnamefont {Y.}~\bibnamefont
  {Castin}},\ }\href {\doibase 10.1103/PhysRevA.76.013613} {\bibfield
  {journal} {\bibinfo  {journal} {Phys. Rev. A}\ }\textbf {\bibinfo {volume}
  {76}},\ \bibinfo {pages} {013613} (\bibinfo {year} {2007})}\BibitemShut
  {NoStop}%
\bibitem [{\citenamefont {Holzmann}\ and\ \citenamefont
  {Krauth}(2008)}]{PhysRevLett.100.190402}%
  \BibitemOpen
  \bibfield  {author} {\bibinfo {author} {\bibfnamefont {M.}~\bibnamefont
  {Holzmann}}\ and\ \bibinfo {author} {\bibfnamefont {W.}~\bibnamefont
  {Krauth}},\ }\href {\doibase 10.1103/PhysRevLett.100.190402} {\bibfield
  {journal} {\bibinfo  {journal} {Phys. Rev. Lett.}\ }\textbf {\bibinfo
  {volume} {100}},\ \bibinfo {pages} {190402} (\bibinfo {year}
  {2008})}\BibitemShut {NoStop}%
\bibitem [{\citenamefont {Simula}\ and\ \citenamefont
  {Blakie}(2006)}]{PhysRevLett.96.020404}%
  \BibitemOpen
  \bibfield  {author} {\bibinfo {author} {\bibfnamefont {T.~P.}\ \bibnamefont
  {Simula}}\ and\ \bibinfo {author} {\bibfnamefont {P.~B.}\ \bibnamefont
  {Blakie}},\ }\href {\doibase 10.1103/PhysRevLett.96.020404} {\bibfield
  {journal} {\bibinfo  {journal} {Phys. Rev. Lett.}\ }\textbf {\bibinfo
  {volume} {96}},\ \bibinfo {pages} {020404} (\bibinfo {year}
  {2006})}\BibitemShut {NoStop}%
\bibitem [{\citenamefont {Simula}\ \emph {et~al.}(2008)\citenamefont {Simula},
  \citenamefont {Davis},\ and\ \citenamefont {Blakie}}]{PhysRevA.77.023618}%
  \BibitemOpen
  \bibfield  {author} {\bibinfo {author} {\bibfnamefont {T.~P.}\ \bibnamefont
  {Simula}}, \bibinfo {author} {\bibfnamefont {M.~J.}\ \bibnamefont {Davis}}, \
  and\ \bibinfo {author} {\bibfnamefont {P.~B.}\ \bibnamefont {Blakie}},\
  }\href {\doibase 10.1103/PhysRevA.77.023618} {\bibfield  {journal} {\bibinfo
  {journal} {Phys. Rev. A}\ }\textbf {\bibinfo {volume} {77}},\ \bibinfo
  {pages} {023618} (\bibinfo {year} {2008})}\BibitemShut {NoStop}%
\bibitem [{\citenamefont {Bisset}\ \emph {et~al.}(2009)\citenamefont {Bisset},
  \citenamefont {Davis}, \citenamefont {Simula},\ and\ \citenamefont
  {Blakie}}]{PhysRevA.79.033626}%
  \BibitemOpen
  \bibfield  {author} {\bibinfo {author} {\bibfnamefont {R.~N.}\ \bibnamefont
  {Bisset}}, \bibinfo {author} {\bibfnamefont {M.~J.}\ \bibnamefont {Davis}},
  \bibinfo {author} {\bibfnamefont {T.~P.}\ \bibnamefont {Simula}}, \ and\
  \bibinfo {author} {\bibfnamefont {P.~B.}\ \bibnamefont {Blakie}},\ }\href
  {\doibase 10.1103/PhysRevA.79.033626} {\bibfield  {journal} {\bibinfo
  {journal} {Phys. Rev. A}\ }\textbf {\bibinfo {volume} {79}},\ \bibinfo
  {pages} {033626} (\bibinfo {year} {2009})}\BibitemShut {NoStop}%
\bibitem [{\citenamefont {Foster}\ \emph {et~al.}(2010)\citenamefont {Foster},
  \citenamefont {Blakie},\ and\ \citenamefont {Davis}}]{PhysRevA.81.023623}%
  \BibitemOpen
  \bibfield  {author} {\bibinfo {author} {\bibfnamefont {C.~J.}\ \bibnamefont
  {Foster}}, \bibinfo {author} {\bibfnamefont {P.~B.}\ \bibnamefont {Blakie}},
  \ and\ \bibinfo {author} {\bibfnamefont {M.~J.}\ \bibnamefont {Davis}},\
  }\href {\doibase 10.1103/PhysRevA.81.023623} {\bibfield  {journal} {\bibinfo
  {journal} {Phys. Rev. A}\ }\textbf {\bibinfo {volume} {81}},\ \bibinfo
  {pages} {023623} (\bibinfo {year} {2010})}\BibitemShut {NoStop}%
\bibitem [{\citenamefont {Nazarenko}(2011)}]{nazarenko2011}%
  \BibitemOpen
  \bibfield  {author} {\bibinfo {author} {\bibfnamefont {S.}~\bibnamefont
  {Nazarenko}},\ }\href@noop {} {\emph {\bibinfo {title} {Wave turbulence}}}\
  (\bibinfo  {publisher} {Springer-Verlag},\ \bibinfo {year}
  {2011})\BibitemShut {NoStop}%
\bibitem [{\citenamefont {Dyachenko}\ \emph {et~al.}(1992)\citenamefont
  {Dyachenko}, \citenamefont {Newell}, \citenamefont {Pushkarev},\ and\
  \citenamefont {Zakharov}}]{DNPZ-bec}%
  \BibitemOpen
  \bibfield  {author} {\bibinfo {author} {\bibfnamefont {A.}~\bibnamefont
  {Dyachenko}}, \bibinfo {author} {\bibfnamefont {A.}~\bibnamefont {Newell}},
  \bibinfo {author} {\bibfnamefont {A.}~\bibnamefont {Pushkarev}}, \ and\
  \bibinfo {author} {\bibfnamefont {V.}~\bibnamefont {Zakharov}},\ }\href@noop
  {} {\bibfield  {journal} {\bibinfo  {journal} {Physica D}\ }\textbf {\bibinfo
  {volume} {57}},\ \bibinfo {pages} {96} (\bibinfo {year} {1992})}\BibitemShut
  {NoStop}%
\bibitem [{\citenamefont {Zakharov}\ and\ \citenamefont
  {Nazarenko}(2005)}]{Zakharov2005203}%
  \BibitemOpen
  \bibfield  {author} {\bibinfo {author} {\bibfnamefont {V.~E.}\ \bibnamefont
  {Zakharov}}\ and\ \bibinfo {author} {\bibfnamefont {S.~V.}\ \bibnamefont
  {Nazarenko}},\ }\href {\doibase 10.1016/j.physd.2004.11.017} {\bibfield
  {journal} {\bibinfo  {journal} {Physica D: Nonlinear Phenomena}\ }\textbf
  {\bibinfo {volume} {201}},\ \bibinfo {pages} {203 } (\bibinfo {year}
  {2005})}\BibitemShut {NoStop}%
\bibitem [{\citenamefont {Lvov}\ \emph {et~al.}(2003)\citenamefont {Lvov},
  \citenamefont {Nazarenko},\ and\ \citenamefont {West}}]{Lvov2003333}%
  \BibitemOpen
  \bibfield  {author} {\bibinfo {author} {\bibfnamefont {Y.}~\bibnamefont
  {Lvov}}, \bibinfo {author} {\bibfnamefont {S.}~\bibnamefont {Nazarenko}}, \
  and\ \bibinfo {author} {\bibfnamefont {R.}~\bibnamefont {West}},\ }\href
  {\doibase 10.1016/S0167-2789(03)00239-2} {\bibfield  {journal} {\bibinfo
  {journal} {Physica D: Nonlinear Phenomena}\ }\textbf {\bibinfo {volume}
  {184}},\ \bibinfo {pages} {333 } (\bibinfo {year} {2003})}\BibitemShut
  {NoStop}%
\bibitem [{\citenamefont {Nazarenko}\ and\ \citenamefont
  {Onorato}(2006)}]{Nazarenko20061}%
  \BibitemOpen
  \bibfield  {author} {\bibinfo {author} {\bibfnamefont {S.}~\bibnamefont
  {Nazarenko}}\ and\ \bibinfo {author} {\bibfnamefont {M.}~\bibnamefont
  {Onorato}},\ }\href {\doibase 10.1016/j.physd.2006.05.007} {\bibfield
  {journal} {\bibinfo  {journal} {Physica D: Nonlinear Phenomena}\ }\textbf
  {\bibinfo {volume} {219}},\ \bibinfo {pages} {1 } (\bibinfo {year}
  {2006})}\BibitemShut {NoStop}%
\bibitem [{\citenamefont {Nazarenko}\ and\ \citenamefont
  {Onorato}(2007)}]{nazon2007}%
  \BibitemOpen
  \bibfield  {author} {\bibinfo {author} {\bibfnamefont {S.}~\bibnamefont
  {Nazarenko}}\ and\ \bibinfo {author} {\bibfnamefont {M.}~\bibnamefont
  {Onorato}},\ }\href {\doibase 10.1007/s10909-006-9271-z} {\bibfield
  {journal} {\bibinfo  {journal} {Journal of Low Temperature Physics}\ }\textbf
  {\bibinfo {volume} {146}},\ \bibinfo {pages} {31} (\bibinfo {year}
  {2007})}\BibitemShut {NoStop}%
\bibitem [{\citenamefont {Connaughton}\ \emph {et~al.}(2005)\citenamefont
  {Connaughton}, \citenamefont {Josserand}, \citenamefont {Picozzi},
  \citenamefont {Pomeau},\ and\ \citenamefont {Rica}}]{connaughton2005}%
  \BibitemOpen
  \bibfield  {author} {\bibinfo {author} {\bibfnamefont {C.}~\bibnamefont
  {Connaughton}}, \bibinfo {author} {\bibfnamefont {C.}~\bibnamefont
  {Josserand}}, \bibinfo {author} {\bibfnamefont {A.}~\bibnamefont {Picozzi}},
  \bibinfo {author} {\bibfnamefont {Y.}~\bibnamefont {Pomeau}}, \ and\ \bibinfo
  {author} {\bibfnamefont {S.}~\bibnamefont {Rica}},\ }\href {\doibase
  10.1103/PhysRevLett.95.263901} {\bibfield  {journal} {\bibinfo  {journal}
  {Phys. Rev. Lett.}\ }\textbf {\bibinfo {volume} {95}},\ \bibinfo {pages}
  {263901} (\bibinfo {year} {2005})}\BibitemShut {NoStop}%
\bibitem [{\citenamefont {D{\"u}ring}\ \emph {et~al.}(2009)\citenamefont
  {D{\"u}ring}, \citenamefont {Picozzi},\ and\ \citenamefont
  {Rica}}]{during2009breakdown}%
  \BibitemOpen
  \bibfield  {author} {\bibinfo {author} {\bibfnamefont {G.}~\bibnamefont
  {D{\"u}ring}}, \bibinfo {author} {\bibfnamefont {A.}~\bibnamefont {Picozzi}},
  \ and\ \bibinfo {author} {\bibfnamefont {S.}~\bibnamefont {Rica}},\
  }\href@noop {} {\bibfield  {journal} {\bibinfo  {journal} {Physica D:
  Nonlinear Phenomena}\ }\textbf {\bibinfo {volume} {238}},\ \bibinfo {pages}
  {1524} (\bibinfo {year} {2009})}\BibitemShut {NoStop}%
\bibitem [{\citenamefont {Vladimirova}\ \emph {et~al.}(2012)\citenamefont
  {Vladimirova}, \citenamefont {Derevyanko},\ and\ \citenamefont
  {Falkovich}}]{Vladimirova}%
  \BibitemOpen
  \bibfield  {author} {\bibinfo {author} {\bibfnamefont {N.}~\bibnamefont
  {Vladimirova}}, \bibinfo {author} {\bibfnamefont {S.}~\bibnamefont
  {Derevyanko}}, \ and\ \bibinfo {author} {\bibfnamefont {G.}~\bibnamefont
  {Falkovich}},\ }\href {\doibase 10.1103/PhysRevE.85.010101} {\bibfield
  {journal} {\bibinfo  {journal} {Phys. Rev. E}\ }\textbf {\bibinfo {volume}
  {85}},\ \bibinfo {pages} {010101} (\bibinfo {year} {2012})}\BibitemShut
  {NoStop}%
\bibitem [{\citenamefont {Berloff}\ and\ \citenamefont
  {Svistunov}(2002)}]{berloff2002ssn}%
  \BibitemOpen
  \bibfield  {author} {\bibinfo {author} {\bibfnamefont {N.}~\bibnamefont
  {Berloff}}\ and\ \bibinfo {author} {\bibfnamefont {B.}~\bibnamefont
  {Svistunov}},\ }\href@noop {} {\bibfield  {journal} {\bibinfo  {journal}
  {Physical Review A}\ }\textbf {\bibinfo {volume} {66}},\ \bibinfo {pages}
  {13603} (\bibinfo {year} {2002})}\BibitemShut {NoStop}%
\bibitem [{\citenamefont {Proment}\ \emph {et~al.}(2012)\citenamefont
  {Proment}, \citenamefont {Nazarenko},\ and\ \citenamefont
  {Onorato}}]{Proment:2012rt}%
  \BibitemOpen
  \bibfield  {author} {\bibinfo {author} {\bibfnamefont {D.}~\bibnamefont
  {Proment}}, \bibinfo {author} {\bibfnamefont {S.}~\bibnamefont {Nazarenko}},
  \ and\ \bibinfo {author} {\bibfnamefont {M.}~\bibnamefont {Onorato}},\
  }\bibfield  {booktitle} {\emph {\bibinfo {booktitle} {Special Issue on Small
  Scale Turbulence}},\ }\href
  {http://www.sciencedirect.com/science/article/pii/S0167278911001564}
  {\bibfield  {journal} {\bibinfo  {journal} {Physica D: Nonlinear Phenomena}\
  }\textbf {\bibinfo {volume} {241}},\ \bibinfo {pages} {304} (\bibinfo {year}
  {2012})}\BibitemShut {NoStop}%
\bibitem [{\citenamefont {Krstulovic}\ and\ \citenamefont
  {Brachet}(2011)}]{PhysRevE.83.066311}%
  \BibitemOpen
  \bibfield  {author} {\bibinfo {author} {\bibfnamefont {G.}~\bibnamefont
  {Krstulovic}}\ and\ \bibinfo {author} {\bibfnamefont {M.}~\bibnamefont
  {Brachet}},\ }\href {\doibase 10.1103/PhysRevE.83.066311} {\bibfield
  {journal} {\bibinfo  {journal} {Phys. Rev. E}\ }\textbf {\bibinfo {volume}
  {83}},\ \bibinfo {pages} {066311} (\bibinfo {year} {2011})}\BibitemShut
  {NoStop}%
\bibitem [{\citenamefont {Schole}\ \emph {et~al.}(2012)\citenamefont {Schole},
  \citenamefont {Nowak},\ and\ \citenamefont {Gasenzer}}]{PhysRevA.86.013624}%
  \BibitemOpen
  \bibfield  {author} {\bibinfo {author} {\bibfnamefont {J.}~\bibnamefont
  {Schole}}, \bibinfo {author} {\bibfnamefont {B.}~\bibnamefont {Nowak}}, \
  and\ \bibinfo {author} {\bibfnamefont {T.}~\bibnamefont {Gasenzer}},\ }\href
  {\doibase 10.1103/PhysRevA.86.013624} {\bibfield  {journal} {\bibinfo
  {journal} {Phys. Rev. A}\ }\textbf {\bibinfo {volume} {86}},\ \bibinfo
  {pages} {013624} (\bibinfo {year} {2012})}\BibitemShut {NoStop}%
\bibitem [{\citenamefont {Sun}\ \emph {et~al.}(2012)\citenamefont {Sun},
  \citenamefont {Jia}, \citenamefont {Barsi}, \citenamefont {Rica},
  \citenamefont {Picozzi},\ and\ \citenamefont
  {Fleischer}}]{sun2012observation}%
  \BibitemOpen
  \bibfield  {author} {\bibinfo {author} {\bibfnamefont {C.}~\bibnamefont
  {Sun}}, \bibinfo {author} {\bibfnamefont {S.}~\bibnamefont {Jia}}, \bibinfo
  {author} {\bibfnamefont {C.}~\bibnamefont {Barsi}}, \bibinfo {author}
  {\bibfnamefont {S.}~\bibnamefont {Rica}}, \bibinfo {author} {\bibfnamefont
  {A.}~\bibnamefont {Picozzi}}, \ and\ \bibinfo {author} {\bibfnamefont
  {J.~W.}\ \bibnamefont {Fleischer}},\ }\href@noop {} {\bibfield  {journal}
  {\bibinfo  {journal} {Nature Physics}\ }\textbf {\bibinfo {volume} {8}},\
  \bibinfo {pages} {471} (\bibinfo {year} {2012})}\BibitemShut {NoStop}%
\bibitem [{\citenamefont {{Shukla}}\ \emph {et~al.}(2013)\citenamefont
  {{Shukla}}, \citenamefont {{Brachet}},\ and\ \citenamefont
  {{Pandit}}}]{Shukla}%
  \BibitemOpen
  \bibfield  {author} {\bibinfo {author} {\bibfnamefont {V.}~\bibnamefont
  {{Shukla}}}, \bibinfo {author} {\bibfnamefont {M.}~\bibnamefont {{Brachet}}},
  \ and\ \bibinfo {author} {\bibfnamefont {R.}~\bibnamefont {{Pandit}}},\
  }\href@noop {} {\bibfield  {journal} {\bibinfo  {journal} {ArXiv e-prints}\ }
  (\bibinfo {year} {2013})},\ \Eprint {http://arxiv.org/abs/1301.3383}
  {arXiv:1301.3383 [nlin.CD]} \BibitemShut {NoStop}%
\bibitem [{\citenamefont {Hadzibabic}\ and\ \citenamefont
  {Dalibard}(2011)}]{Hadzibabic}%
  \BibitemOpen
  \bibfield  {author} {\bibinfo {author} {\bibfnamefont {Z.}~\bibnamefont
  {Hadzibabic}}\ and\ \bibinfo {author} {\bibfnamefont {J.}~\bibnamefont
  {Dalibard}},\ }\href@noop {} {\bibfield  {journal} {\bibinfo  {journal}
  {Rivista del Nuovo Cimento}\ }\textbf {\bibinfo {volume} {34}},\ \bibinfo
  {pages} {389} (\bibinfo {year} {2011})}\BibitemShut {NoStop}%
\bibitem [{\citenamefont {Kogut}(1979)}]{RevModPhys.51.659}%
  \BibitemOpen
  \bibfield  {author} {\bibinfo {author} {\bibfnamefont {J.~B.}\ \bibnamefont
  {Kogut}},\ }\href {\doibase 10.1103/RevModPhys.51.659} {\bibfield  {journal}
  {\bibinfo  {journal} {Rev. Mod. Phys.}\ }\textbf {\bibinfo {volume} {51}},\
  \bibinfo {pages} {659} (\bibinfo {year} {1979})}\BibitemShut {NoStop}%
\bibitem [{\citenamefont {Pierson}(1997)}]{doi:10.1080/01418639708241138}%
  \BibitemOpen
  \bibfield  {author} {\bibinfo {author} {\bibfnamefont {S.~W.}\ \bibnamefont
  {Pierson}},\ }\href {\doibase 10.1080/01418639708241138} {\bibfield
  {journal} {\bibinfo  {journal} {Philosophical Magazine Part B}\ }\textbf
  {\bibinfo {volume} {76}},\ \bibinfo {pages} {715} (\bibinfo {year} {1997})},\
  \Eprint
  {http://arxiv.org/abs/http://www.tandfonline.com/doi/pdf/10.1080/01418639708241138}
  {http://www.tandfonline.com/doi/pdf/10.1080/01418639708241138} \BibitemShut
  {NoStop}%
\bibitem [{\citenamefont {Kozik}\ and\ \citenamefont
  {Svistunov}(2005)}]{PhysRevB.72.172505}%
  \BibitemOpen
  \bibfield  {author} {\bibinfo {author} {\bibfnamefont {E.}~\bibnamefont
  {Kozik}}\ and\ \bibinfo {author} {\bibfnamefont {B.}~\bibnamefont
  {Svistunov}},\ }\href {\doibase 10.1103/PhysRevB.72.172505} {\bibfield
  {journal} {\bibinfo  {journal} {Phys. Rev. B}\ }\textbf {\bibinfo {volume}
  {72}},\ \bibinfo {pages} {172505} (\bibinfo {year} {2005})}\BibitemShut
  {NoStop}%
\bibitem [{\citenamefont {Cichowlas}\ \emph {et~al.}(2005)\citenamefont
  {Cichowlas}, \citenamefont {Bona{\"\i}ti}, \citenamefont {Debbasch},\ and\
  \citenamefont {Brachet}}]{PhysRevLett.95.264502}%
  \BibitemOpen
  \bibfield  {author} {\bibinfo {author} {\bibfnamefont {C.}~\bibnamefont
  {Cichowlas}}, \bibinfo {author} {\bibfnamefont {P.}~\bibnamefont
  {Bona{\"\i}ti}}, \bibinfo {author} {\bibfnamefont {F.}~\bibnamefont
  {Debbasch}}, \ and\ \bibinfo {author} {\bibfnamefont {M.}~\bibnamefont
  {Brachet}},\ }\href {\doibase 10.1103/PhysRevLett.95.264502} {\bibfield
  {journal} {\bibinfo  {journal} {Phys. Rev. Lett.}\ }\textbf {\bibinfo
  {volume} {95}},\ \bibinfo {pages} {264502} (\bibinfo {year}
  {2005})}\BibitemShut {NoStop}%
\bibitem [{\citenamefont {Ray}\ \emph {et~al.}(2011)\citenamefont {Ray},
  \citenamefont {Frisch}, \citenamefont {Nazarenko},\ and\ \citenamefont
  {Matsumoto}}]{PhysRevE.84.016301}%
  \BibitemOpen
  \bibfield  {author} {\bibinfo {author} {\bibfnamefont {S.~S.}\ \bibnamefont
  {Ray}}, \bibinfo {author} {\bibfnamefont {U.}~\bibnamefont {Frisch}},
  \bibinfo {author} {\bibfnamefont {S.}~\bibnamefont {Nazarenko}}, \ and\
  \bibinfo {author} {\bibfnamefont {T.}~\bibnamefont {Matsumoto}},\ }\href
  {\doibase 10.1103/PhysRevE.84.016301} {\bibfield  {journal} {\bibinfo
  {journal} {Phys. Rev. E}\ }\textbf {\bibinfo {volume} {84}},\ \bibinfo
  {pages} {016301} (\bibinfo {year} {2011})}\BibitemShut {NoStop}%
\bibitem [{\citenamefont {Davis}\ \emph {et~al.}(2001)\citenamefont {Davis},
  \citenamefont {Morgan},\ and\ \citenamefont
  {Burnett}}]{PhysRevLett.87.160402}%
  \BibitemOpen
  \bibfield  {author} {\bibinfo {author} {\bibfnamefont {M.~J.}\ \bibnamefont
  {Davis}}, \bibinfo {author} {\bibfnamefont {S.~A.}\ \bibnamefont {Morgan}}, \
  and\ \bibinfo {author} {\bibfnamefont {K.}~\bibnamefont {Burnett}},\ }\href
  {\doibase 10.1103/PhysRevLett.87.160402} {\bibfield  {journal} {\bibinfo
  {journal} {Phys. Rev. Lett.}\ }\textbf {\bibinfo {volume} {87}},\ \bibinfo
  {pages} {160402} (\bibinfo {year} {2001})}\BibitemShut {NoStop}%
\bibitem [{\citenamefont {Newell}\ and\ \citenamefont
  {Moloney}(1990)}]{newell1990nonlinear}%
  \BibitemOpen
  \bibfield  {author} {\bibinfo {author} {\bibfnamefont {A.~C.}\ \bibnamefont
  {Newell}}\ and\ \bibinfo {author} {\bibfnamefont {J.}~\bibnamefont
  {Moloney}},\ }in\ \href@noop {} {\emph {\bibinfo {booktitle} {Partially
  Intergrable Evolution Equations in Physics}}}\ (\bibinfo  {publisher}
  {Springer},\ \bibinfo {year} {1990})\ pp.\ \bibinfo {pages}
  {161--161}\BibitemShut {NoStop}%
\bibitem [{\citenamefont {Kivshar}\ and\ \citenamefont
  {Agrawal}(2003)}]{kivshar2003optical}%
  \BibitemOpen
  \bibfield  {author} {\bibinfo {author} {\bibfnamefont {Y.~S.}\ \bibnamefont
  {Kivshar}}\ and\ \bibinfo {author} {\bibfnamefont {G.}~\bibnamefont
  {Agrawal}},\ }\href@noop {} {\emph {\bibinfo {title} {Optical solitons: from
  fibers to photonic crystals}}}\ (\bibinfo  {publisher} {Academic press},\
  \bibinfo {year} {2003})\BibitemShut {NoStop}%
\bibitem [{\citenamefont {Boyd}(1992)}]{boyd1992nonlinear}%
  \BibitemOpen
  \bibfield  {author} {\bibinfo {author} {\bibfnamefont {R.~W.}\ \bibnamefont
  {Boyd}},\ }\href@noop {} {\  (\bibinfo {year} {1992})}\BibitemShut {NoStop}%
\bibitem [{\citenamefont {Picozzi}\ and\ \citenamefont
  {Rica}(2012)}]{Picozzi20125440}%
  \BibitemOpen
  \bibfield  {author} {\bibinfo {author} {\bibfnamefont {A.}~\bibnamefont
  {Picozzi}}\ and\ \bibinfo {author} {\bibfnamefont {S.}~\bibnamefont {Rica}},\
  }\href {\doibase http://dx.doi.org/10.1016/j.optcom.2012.07.081} {\bibfield
  {journal} {\bibinfo  {journal} {Optics Communications}\ }\textbf {\bibinfo
  {volume} {285}},\ \bibinfo {pages} {5440 } (\bibinfo {year}
  {2012})}\BibitemShut {NoStop}%
\bibitem [{\citenamefont {{Suret}}\ and\ \citenamefont
  {{Randoux}}(2013)}]{2013arXiv1307.5034S}%
  \BibitemOpen
  \bibfield  {author} {\bibinfo {author} {\bibfnamefont {P.}~\bibnamefont
  {{Suret}}}\ and\ \bibinfo {author} {\bibfnamefont {S.}~\bibnamefont
  {{Randoux}}},\ }\href@noop {} {\bibfield  {journal} {\bibinfo  {journal}
  {ArXiv e-prints}\ } (\bibinfo {year} {2013})},\ \Eprint
  {http://arxiv.org/abs/1307.5034} {arXiv:1307.5034 [physics.optics]}
  \BibitemShut {NoStop}%
\bibitem [{\citenamefont {Pomeau}(1992)}]{0951-7715-5-3-005}%
  \BibitemOpen
  \bibfield  {author} {\bibinfo {author} {\bibfnamefont {Y.}~\bibnamefont
  {Pomeau}},\ }\href {http://stacks.iop.org/0951-7715/5/i=3/a=005} {\bibfield
  {journal} {\bibinfo  {journal} {Nonlinearity}\ }\textbf {\bibinfo {volume}
  {5}},\ \bibinfo {pages} {707} (\bibinfo {year} {1992})}\BibitemShut {NoStop}%
\bibitem [{Note1()}]{Note1}%
  \BibitemOpen
  \bibinfo {note} {Note that this point is particularly important when
  comparing our defined non-dimensional 2D healing length with the one usually
  defined in the BEC community. First, one should recall that our definition is
  for a two-dimensional NLS equation and so appropriate scalings should be used
  as shown in Appendix \ref {app:NLS-BEC}. Second, our definition depends on
  the total number of bosons in the system, not only on the macroscopic number
  present in the condensate: we recover the usual definition only if localised
  perturbations are negligible with respect of the uniform condensate
  background.}\BibitemShut {Stop}%
\bibitem [{\citenamefont {Zakharov}\ \emph {et~al.}(1985)\citenamefont
  {Zakharov}, \citenamefont {Musher},\ and\ \citenamefont
  {Rubenchik}}]{musher-nls}%
  \BibitemOpen
  \bibfield  {author} {\bibinfo {author} {\bibfnamefont {V.}~\bibnamefont
  {Zakharov}}, \bibinfo {author} {\bibfnamefont {S.}~\bibnamefont {Musher}}, \
  and\ \bibinfo {author} {\bibfnamefont {A.}~\bibnamefont {Rubenchik}},\
  }\href@noop {} {\bibfield  {journal} {\bibinfo  {journal} {Phys. Rep.}\
  }\textbf {\bibinfo {volume} {129}},\ \bibinfo {pages} {285} (\bibinfo {year}
  {1985})}\BibitemShut {NoStop}%
\bibitem [{\citenamefont {Zakharov}\ \emph {et~al.}(1992)\citenamefont
  {Zakharov}, \citenamefont {L'vov},\ and\ \citenamefont
  {Falkovich}}]{zakharov41kst}%
  \BibitemOpen
  \bibfield  {author} {\bibinfo {author} {\bibfnamefont {V.}~\bibnamefont
  {Zakharov}}, \bibinfo {author} {\bibfnamefont {V.}~\bibnamefont {L'vov}}, \
  and\ \bibinfo {author} {\bibfnamefont {G.}~\bibnamefont {Falkovich}},\
  }\href@noop {} {\emph {\bibinfo {title} {{Kolmogorov Spectra of Turbulence 1:
  Wave Turbulence}}}}\ (\bibinfo  {publisher} {Springer-Verlag},\ \bibinfo
  {year} {1992})\BibitemShut {NoStop}%
\bibitem [{\citenamefont {Popov}(1983)}]{Popov}%
  \BibitemOpen
  \bibfield  {author} {\bibinfo {author} {\bibfnamefont {V.}~\bibnamefont
  {Popov}},\ }\href@noop {} {\emph {\bibinfo {title} {Functional Integrals in
  Quantum Field Theory and Statistical Physics}}}\ (\bibinfo  {publisher}
  {Reidel, Dordrecht},\ \bibinfo {year} {1983})\BibitemShut {NoStop}%
\bibitem [{\citenamefont {Holzmann}\ and\ \citenamefont
  {Baym}(2007)}]{PhysRevB.76.092502}%
  \BibitemOpen
  \bibfield  {author} {\bibinfo {author} {\bibfnamefont {M.}~\bibnamefont
  {Holzmann}}\ and\ \bibinfo {author} {\bibfnamefont {G.}~\bibnamefont
  {Baym}},\ }\href {\doibase 10.1103/PhysRevB.76.092502} {\bibfield  {journal}
  {\bibinfo  {journal} {Phys. Rev. B}\ }\textbf {\bibinfo {volume} {76}},\
  \bibinfo {pages} {092502} (\bibinfo {year} {2007})}\BibitemShut {NoStop}%
\bibitem [{\citenamefont {Holzmann}\ \emph {et~al.}(2007)\citenamefont
  {Holzmann}, \citenamefont {Baym}, \citenamefont {Blaizot},\ and\
  \citenamefont {Lalo{\"e}}}]{Holzmann30012007}%
  \BibitemOpen
  \bibfield  {author} {\bibinfo {author} {\bibfnamefont {M.}~\bibnamefont
  {Holzmann}}, \bibinfo {author} {\bibfnamefont {G.}~\bibnamefont {Baym}},
  \bibinfo {author} {\bibfnamefont {J.-P.}\ \bibnamefont {Blaizot}}, \ and\
  \bibinfo {author} {\bibfnamefont {F.}~\bibnamefont {Lalo{\"e}}},\ }\href
  {http://www.pnas.org/content/104/5/1476.abstract N2 - Current experiments on
  atomic gases in highly anisotropic traps present the opportunity to study in
  detail the low temperature phases of two-dimensional inhomogeneous systems.
  Although, in an ideal gas, the trapping potential favors Bose–Einstein
  condensation at finite temperature, interactions tend to destabilize the
  condensate, leading to a superfluid Kosterlitz–Thouless–Berezinskii phase
  with a finite superfluid mass density but no long-range order, as in
  homogeneous fluids. The transition in homogeneous systems is conveniently
  described in terms of dissociation of topological defects
  (vortex–antivortex pairs). However, trapped two-dimensional gases are more
  directly approached by generalizing the microscopic theory of the homogeneous
  gas. In this paper, we first derive, via a diagrammatic expansion, the
  scaling structure near the phase transition in a homogeneous system, and then
  study the effects of a trapping potential in the local density approximation.
  We find that a weakly interacting trapped gas undergoes a
  Kosterlitz–Thouless–Berezinskii transition from the normal state at a
  temperature slightly below the Bose–Einstein transition temperature of the
  ideal gas. The characteristic finite superfluid mass density of a homogeneous
  system just below the transition becomes strongly suppressed in a trapped
  gas.} {\bibfield  {journal} {\bibinfo  {journal} {Proceedings of the National
  Academy of Sciences}\ }\textbf {\bibinfo {volume} {104}},\ \bibinfo {pages}
  {1476} (\bibinfo {year} {2007})}\BibitemShut {NoStop}%
\bibitem [{\citenamefont {Nazarenko}\ and\ \citenamefont {West}(2003)}]{west}%
  \BibitemOpen
  \bibfield  {author} {\bibinfo {author} {\bibfnamefont {S.}~\bibnamefont
  {Nazarenko}}\ and\ \bibinfo {author} {\bibfnamefont {R.}~\bibnamefont
  {West}},\ }\href {\doibase 10.1023/A:1023719007403} {\bibfield  {journal}
  {\bibinfo  {journal} {Journal of Low Temperature Physics}\ }\textbf {\bibinfo
  {volume} {132}},\ \bibinfo {pages} {1} (\bibinfo {year} {2003})}\BibitemShut
  {NoStop}%
\end{thebibliography}%

\end{document}